\documentclass[%
reprint,
 aps,
 prb,
 longbibliography,
]{revtex4-2}

\usepackage{amsmath,amssymb}
\usepackage[dvipdfmx]{graphicx}

\usepackage{braket}
\usepackage{here}
\usepackage{comment}
\usepackage{color}
\usepackage{diagbox} 
\usepackage{colortbl}
\usepackage{dcolumn}
\usepackage{bbm}
\usepackage{bm}
\usepackage{ulem}
\usepackage{lipsum}

\usepackage[dvipsnames]{xcolor}

\def\rr{{\bm r}}

\def\kk{{\bm k}}

\def\VX{V^{1\mathrm{X}}_\mathrm{rms}}
\def\VY{V^{2\mathrm{Y}}_\mathrm{rms}}
\def\RY{\overline{R}^{2\mathrm{Y}}_{zz}}
\def\Irms{I_\mathrm{rms}}
\def\gzz{g_{zz}}
\def\Gzz{G_{zz}}
\def\gzzzz{g_{zzzz}}
\def\Gzzzz{G_{zzzz}}
\def\Order{\mathcal{O}}

\def\Integ{\mathcal{I}}
\def\dkf{\frac{\partial f}{\partial \bm{k}}}

\def\Ev{\widetilde{E}_\mathrm{v}}

\begin{document}
\preprint{APS/123-QED}
\title{Chiral orbital texture in nonlinear electrical conduction}

    \author{Suguru Okumura}
    \affiliation{Shizuoka University, Suruga, Shizuoka, 422-8529, Japan.}

    \author{Ryutaro Tanaka}
    \affiliation{Shizuoka University, Suruga, Shizuoka, 422-8529, Japan.}

    \author{Daichi Hirobe}
    \affiliation{Shizuoka University, Suruga, Shizuoka, 422-8529, Japan.}

\begin{abstract}
Nonlinear electrical conduction primarily mediated by an orbital texture is observed in chiral semiconductor Te. 
We determine the enantiospecific sign of the nonlinear conductance and identify anomalies in its carrier-density dependence. 
Our findings, combined with the Boltzmann equation, are attributed to a chiral orbital texture, namely a chiral distribution of the orbital magnetic moment in reciprocal space. 
This study underscores the efficacy of nonlinear transport measurements in probing orbital-related effects, whose differentiation from spin counterparts is often demanding in the linear response regime of electron transport.
\end{abstract}
\maketitle

{\it Introduction.}---
Magnetoresistance (MR) denotes a variation in electrical resistance in response to a magnetic field, a phenomenon ubiquitous not only in magnetically ordered conductors but also in nonmagnetic conductors.
In conventional MR effects in nonmagnetic conductors, the resistance change depends solely on the strength of a magnetic field, irrespective of its polarity. However, this paradigm changes when the system breaks inversion symmetry, rendering it noncentrosymmetric.
Such noncentrosymmetry can produce MR that varies bilinearly with both the magnetic field $B$ and the electric current $I$~\cite{RikkenPRL2001, RikkenPRL2005}. 
This bilinear MR produces a voltage proportional to the product of $I^2$ and $B$, underscoring its nonlinearity and unipolar dependence on $B$. 
Due to the characteristics, bilinear MR is also referred to as nonreciprocal magnetotransport~\cite{TokuraNatCom2018}, bulk charge rectification~\cite{ideueNatPhys2017}, magnetochiral anisotropy~\cite{PopNatCommun2014, YokouchiPRL2017}, and unidirectional magnetoresistance~\cite{AvciNatPhys2015}. 
This capability to rectify an electric current without the need for heterojunctions holds promise for utilizing noncentrosymmetric properties in electronics. 
For its direct relevance to our observables, we adopt the term nonlinear electrical conduction (NEC) in this paper to mean bilinear MR in noncentrosymmetric conductors lacking magnetic ordering and heterojunctions.

NEC can arise from several sources, including the Zeeman interaction~\cite{ideueNatPhys2017, HeNatCommun2019} and the Berry curvature~\cite{YokouchiPRL2023}. 
In the context of noncentrosymmetry, the spin texture in quasimomentum ($\bm{k}$) space is often emphasized.
Notably, an orbital counterpart equally represents noncentrosymmetry without the need for spin--orbit coupling: the orbital magnetic moment of a Bloch electron in $\kk$ space~\cite{ChangPhysRevB1996, XiaoRevModPhys2010}. 
The orbital texture can be conceived as the $\bm{k}$-dependent self-rotation of a wavepacket about its center of mass in semiclassical theory.
Theoretical predictions suggest that the orbital texture enables current-induced magnetization~\cite{YodaSciRep2015, ZhongPRL2016, YodaNanoLett2018} and may even dominate over the spin texture in certain scenarios~\cite{HeNatCommun2020, HePhysRevRes2021, ChirolliPRL2022}. 
Recently, a similar theoretical argument has been applied to NEC, focusing on the three-dimensional chiral semiconductor Te~\cite{liu2023electrical}. 
Experimental indication of the relevance of the orbital magnetic moment near the Weyl points of Weyl semimetal WTe$_2$ was provided through the divergent behavior of magnetochiral anisotropy~\cite{YokouchiPRL2023}.
However, exploration of orbital-related nonlinear transport remains largely uncharted, raising the question of whether orbital-induced NEC is universal beyond materials exhibiting such topological singularities.

In this article, we report NEC induced primarily by a chiral orbital texture in p-type Te, where the chemical potential resides near the highest valence band without topological singularities.
We confirm chirality-induced NEC in Te-based field-effect transistors, obeying the magnetic group of trigonal Te. 
Using symmetry-adapted selection rules, we determine a single component $\Gzzzz$ of the nonlinear electrical conductance tensor as a function of carrier density with high precision.
The enantiospecific sign of $\Gzzzz$ is negative (positive) for right-handed (left-handed) Te, and $\Gzzzz$ exhibits a broad peak structure at low carrier densities.
These characteristics are attributed primarily to a chiral orbital texture in $\kk$ space, rather than the spin counterpart. This is rationalized by Boltzmann kinetic theory incorporating orbital and spin textures via the Zeeman interaction. 
Discriminating between orbital and spin contributions is often challenging in dynamical magnetoelectric effects.
Our findings suggest that NEC measurements offer an effective means of probing elusive orbital textures across diverse materials.

{\it Chiral properties of elemental tellurium.}---
Trigonal Te is a p-type semiconductor with a narrow band gap of approximately $0.34$ eV. Helical chains form along the $c$ axis through covalent bonding of Te atoms, with adjacent chains interconnected via coordinate covalent bonding due to the multivalent nature of Te~\cite{ZhuPRL2017, YiInorgChem2018}.
Consequently, Te crystallizes into enantiomorphic space groups, either $P3_{1}21$ (right-handed) or $P3_{2}21$ (left-handed) [Fig.~\ref{fig:chiral_properties}(a)]. 
Numerous phenomena in Te are attributed to chirality, including hedgehoglike spin textures~\cite{SakanoPRL2020, GattiPRL2020}, optical activity~\cite{NomuraPRL1960, FukudaPSS1975, AdesJOSA1975, StolzePSS1977}, current-induced magnetization~\cite{VorobevPETF1979, ShalyginPSS2012, FurukawaNatCom2017, FurukawaPhysRevRes2021}, asymmetric etch pits~\cite{koma1970etch}, NEC~\cite{CalavalleNatMater2022, hirobe2022chirality, sudo2023valley, niu2023tunable00}, second-harmonic generation~\cite{ChengPRB2019, FuAdvMater2023,niu2023tunable00}, circular photogalvanic and photovoltaic effects~\cite{niu2023tunable01}, and diffraction with circularly polarized x rays~\cite{TanakaJPCM2010}.
The simplicity of its chiral structure has facilitated theoretical calculations of chirality-related properties for decades~\cite{MatibetPSS1969, DoiJPSJ1970, joannopoulos1975electronic, asendorf1957space, HirayamaPRL2015, peng2014elemental, ivchenko1978new, TsirkinPRB2018, csahin2018pancharatnam}. For modeling energy dispersions and magnetic moments, we adopt the tight-binding model proposed in Ref.~\onlinecite{csahin2018pancharatnam}, wherein the reduced space is spanned by two conduction bands and four valence bands. Original parameters are adjusted to match {\it ab initio} calculations~\cite{TsirkinPRB2018}, which successfully reproduce experimentally confirmed enantiomeric spin magnetic moments~\cite{SakanoPRL2020,FurukawaPhysRevRes2021}. The energy dispersion of the uppermost valence band near the H and H' points [Fig.~\ref{fig:chiral_properties}(b)] is well approximated by $E_{\mathrm{v}}(\bm{k}) = -\hbar^2/(2m_\perp^\mathrm{v})(k_x^2 + k_y^2) - \hbar^2/(2m_\parallel^\mathrm{v})k_z^2 + \sqrt{(S k_z)^2+\varDelta^2}-\varDelta$ [Fig.~\ref{fig:chiral_properties}(c)],
where $\hbar$ is the reduced Planck constant, and other parameter values are provided in the Supplemental Material~\cite{SM} (see also Refs.~\cite{schliemann2003anisotropic, vyborny2009semiclassical, liu2016mobility, xiao2016unconventional, kim2019vertex} therein).  
The $z$ and $x$ axes are aligned with the $c$ and $a$ axes of Te, respectively. 
The spin magnetic moment $m_z^{\mathrm{spin}}(\bm{k})=\bm{m}^{\mathrm{spin}}(\bm{k})\cdot\bm{e}_z$ ($\bm{e}_z$: unit vector along the $z$ axis) for $E_{\mathrm{v}}(\bm{k})$ is represented by $m_z^{\mathrm{spin}}(\bm{k}) = -\mu_\mathrm{B}\eta(k_z)$,
where $\eta(k_z) = S k_z/\sqrt{(S k_z)^2+\varDelta^2}$ and $\mu_\mathrm{B}$ is the Bohr magneton.
Crucially, the parameter $S$ in $\eta(k_z)$ changes sign with the handedness of the lattice structure.
Thus, $\bm{m}^{\mathrm{spin}}(\bm{k})$ is handedness-dependent and hedgehoglike in $\bm{k}$ space, which is a hallmark of chirality. 

\begin{figure}[ht]
\begin{center}
\includegraphics[width=70mm]{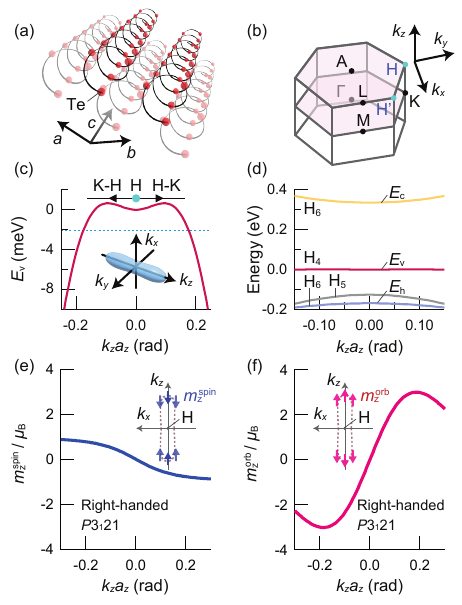}
\end{center}
\caption{
(a) Crystal structure of right-handed Te of space group $P3_121$.
(b) First Brillouin zone and representative highly symmetrical points. 
$k_{x,y,z}$ denote wavenumbers, measured from the H point. 
(c), (d) $k_z$ dependences of the valence band $E_\mathrm{v}$ (c) and other bands located around $E_\mathrm{v}$ (d) in the vicinity of the H point. $a_z$ is the lattice constant along the $c$ axis. The Fermi surface of $E_\mathrm{v}$ at $-2$~meV is shown in (c). 
(e), (f) $k_z$ dependences of the $z$ components of the spin magnetic moment $m_z^\mathrm{spin}$ (e) and the orbital magnetic moment $m_z^\mathrm{orb}$ (f) for right-handed Te ($P$3$_1$21). 
}
\label{fig:chiral_properties}
\end{figure}

Such chirality is also evident in the orbital texture, even in the absence of spin--orbit coupling, in principle.
The orbital magnetic moment $m_z^{\mathrm{orb}}(\kk)=\bm{m}^{\mathrm{orb}}(\kk)\cdot\bm{e}_z=-ie/(2\hbar)\bra{\nabla_\kk u_\mathrm{v}}\times(H_\kk-E_\mathrm{v}(\kk))\ket{\nabla_\kk u_\mathrm{v}}\cdot\bm{e}_z$ ($e>0$: elementary charge) for the highest valence band can be approximated as~\cite{csahin2018pancharatnam}
\begin{equation}
    m_z^{\mathrm{orb}}(\bm{k})
    =
    -\mu_\mathrm{B}\left(
        \frac{\epsilon_\mathrm{c}}{\varDelta E_\mathrm{vc}(\bm{k})} + \frac{\epsilon_\mathrm{v}}{\varDelta E_\mathrm{vh}(\bm{k})}
    \right)
    \eta(k_z)
\end{equation}
with $\varDelta E_{\mathrm{vc(vh)}}(\bm{k})=E_{\mathrm{v}}(\bm{k})-E_{\mathrm{c(h)}}(\bm{k})$ denoting the energy difference between $E_\mathrm{v}$ and the highest conduction (lowest valence) band [see also Fig.~\ref{fig:chiral_properties}(d)]. 
Here, $\epsilon_\mathrm{c}=4.965$~eV and $\epsilon_\mathrm{h}=1.745$~eV are parameters adjusted to match {\it ab initio} calculations~\cite{SM}.
As illustrated in Figs.~1(e) and (f), $m_z^{\mathrm{orb}}$ not only surpasses $m_z^{\mathrm{spin}}$ in magnitude but also exhibits the opposite sign.
Moreover, the energy dependence of $m_z^{\mathrm{orb}}$ differs from that of $m_z^{\mathrm{spin}}$ due to a suppression for higher $\kk$ [Fig. 1(f)], arising from the interband nature of the orbital magnetic moment~\cite{XiaoRevModPhys2010}.
These distinctive features of the orbital texture motivated us to explore the corresponding NEC.

{\it Experimental details.}---
Te slabs were synthesized with several modifications to the original protocol~\cite{WanNatElectron2018, SM}: specifically, we reduced the amount of reducing agent in the hydrothermal synthesis and extended the reaction time approximately threefold to produce thicker Te slabs.
These Te slabs exhibited NEC based on the three-dimensional magnetic group while retaining gate-variable resistances.
The Te slabs were laminated on a $\mathrm{SiO_2}$ dielectric (300 nm thick) atop the doped Si substrate [Fig.~\ref{fig:absolute_sign}(a)].
Electrode patterns were defined using standard photolithography for three devices: 
devices A and B included two-terminal electrodes, while device C included Hall and four-terminal electrodes.
Metal electrodes were deposited via electron beam evaporation, with a 20 nm thick layer of Ni followed by a 60 nm thick layer of Au for capping to prevent Ni oxidation. 
Ni was selected to suppress the energy band bending near the Te/electrode interface~\cite{qiu2019thermoelectric}.
For voltage measurements, harmonic voltages were detected using phase-sensitive detection combined with pulse amplitude modulation to minimize the self-heating of devices.
Electric current pulses were modulated in a regularly timed sequence, resulting in a sinusoidal waveform of the envelope $I(t) = \sqrt{2}\Irms\sin(2\pi ft)$, where $t$ denotes time; $f$ a frequency; $I_\mathrm{rms}$ a root mean square current. $f$ was referenced for phase-sensitive detection of root mean square harmonic voltages $\VX$ at $f$ in phase with $I(t)$ and $\VY$ at $2f$ out of phase with $I(t)$. 
Note that uppercase X and Y denote phase relations in phase-sensitive detection, not Cartesian coordinates.
All measurements were conducted in a custom-made cryostat under a vacuum level of approximately $1\times10^{-5}$ Pa.
Magnetic field up to 500~mT was applied by electromagnet.
The space group of devices except device C was determined by observing asymmetric etch pits formed using hot sulfuric acid [Fig.~\ref{fig:absolute_sign}(b)] before electrical measurements, with reference to Ref.~\onlinecite{koma1970etch}.

\begin{figure}[hb]
\begin{center}
\includegraphics[width=80mm]{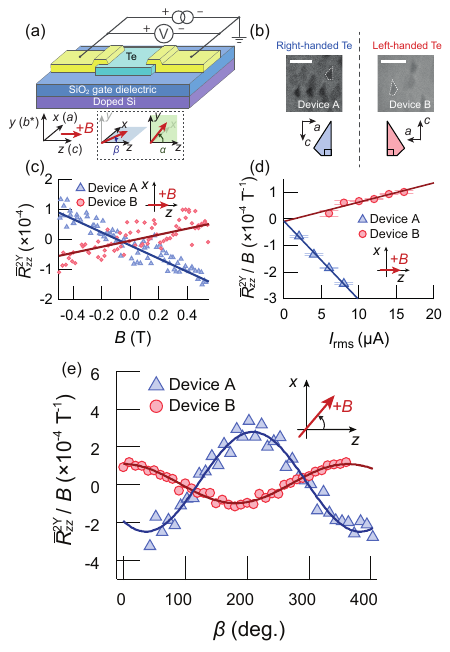}
\end{center}
\caption{
(a) Schematic of the experimental setup for devices A and B. Magnetic field is denoted by $B$, and its angles by $\alpha$ in the $yz$ plane and $\beta$ in the $zx$ plane.
(b) Scanning electron microscope images of asymmetric etch pits formed on the (10$\overline{1}$0) surface. Scale bar: 1~\textmu m.
(c), (d) Bilinear dependence of the normalized second harmonic resistance $\RY$ on $B$ (c) and electric current $I_\mathrm{rms}$ (d), measured at 50~K. 
(e) $\beta$ dependence of $\RY$. $I_\mathrm{rms}$ was set to 10 and 18~\textmu A for devices A and B, respectively, in (c) and (e).
}
\label{fig:absolute_sign}
\end{figure}

\begin{figure*}[ht]
\begin{center}
\includegraphics[width=150mm]{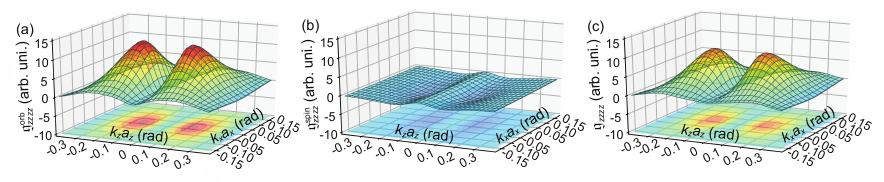}
\end{center}
\caption{
(a)--(c) $\kk$-resolved nonlinear conductivity of left-handed Te for orbital $\mathfrak{g}_{zzzz}^\mathrm{orb}$ (a), spin $\mathfrak{g}_{zzzz}^\mathrm{spin}$ (b) and their sum $\mathfrak{g}_{zzzz}$ (c).
}
\label{fig:g_frak}
\end{figure*}

{\it Absolute sign of the nonlinear electrical conductance.}---
In Figs. \ref{fig:absolute_sign}(c) and (d), we present the normalized second harmonic resistance, $\RY = \VY/\VX$, as a function of the magnetic field $B$ and $\Irms$ at 50~K, both applied along the $z$-axis. 
$\RY$ is $B$- and $\Irms$-linear for both enantiomers, consistent with bilinear MR. Notably, sign reversal of $\RY$ is observed between the two enantiomers.
Further confirmation of chirality-induced NEC is provided by the dependence of the slope $\RY / B$ on the magnetic field angle, $\beta$, measured from the $z$ axis in the $zx$ plane [Fig. \ref{fig:absolute_sign}(e)]. 
$\RY / B$ collapses onto a single cosine wave for each enantiomer and exhibits sign reversal between the two enantiomers over the entire range of $\beta$. 
The observed behavior is fully consistent with the selection rule deduced from the magnetic group of trigonal Te.
For Te thin films and nanowires, the dependence on $\beta$ often deviated from $\cos\beta$ ~\cite{CalavalleNatMater2022,niu2023tunable00}, complicating comparison with theoretical discussions on the origin of NEC. 

We note that the absolute magnitude of $\RY / B$ was not the same between devices A and B at the same excitation current, which could be attributed to different charge rectification efficiencies due to different chemical-potential positions~\cite{hirobe2022chirality,liu2023electrical}. 
We found semiconducting and metallic behaviors for devices A and B, respectively, by temperature-variable harmonic resistance measurement~\cite{SM}. Therefore, device A is expected to exhibit the higher rectification efficiency due to the higher chemical-potential position~\cite{hirobe2022chirality,liu2023electrical}, which is consistent with the higher value of $\RY / B$. In addition, we confirmed that the absolute sign and magnitude of $\RY / B$ was consistent between two-terminal and four-terminal measurements using another device~\cite{SM}, which demonstrates the negligibly small contribution of contact resistance to NEC in our experiment.

Our results allow precise determination of a finite component $\Gzzzz$ of the nonlinear electrical conductance tensor. $\Gzzzz$ appears in a longitudinal electric current $I_z = G_{zz}V_z + G_{zzzz}(V_z)^2B_z + \mathcal{O}((V_z)^3, (B_z)^2)$, where $I_z, V_z$ and $B_z$ denote $z$ components of electric current, voltage and magnetic field while $G_{zz}$ denotes linear electrical conductance. Direct computation yields $\RY = 2^{-1/2}(G_{zz})^{-2}G_{zzzz}\Irms B\cos\beta$~\cite{SM}, enabling determination of $G_{zzzz}$ by fitting to $\RY / B$ with $\cos\beta$. 
We find $G_{zzzz}=-14.5\pm0.6$~nAV$^{-2}$T$^{-1}$ for right-handed Te (device A) and $G_{zzzz}=220\pm8$~nAV$^{-2}$T$^{-1}$ for left-handed Te (device B). The enantiospecific sign was double-checked by d.c. magnetoconductance measurement~\cite{SM}.

We address the enantiospecific sign of $G_{zzzz}$ based on orbital and spin textures, which has not been discussed in previous studies.
We calculate the corresponding nonlinear electrical conductivity $g_{zzzz}$ due to the Zeeman interaction for the energy dispersion $E_\mathrm{v}(\bm{k})-\{m_z^\mathrm{orb}(\bm{k})+m_z^\mathrm{spin}(\bm{k})\}B_z$, following the Boltzmann kinetic theory which accounts for elastic impurity scattering~\cite{golub2023electrical}. 
Because this energy band is separated from the second highest valence band and the lowest conduction band by about 1,500~K and 4,000~K at the H point, respectively, electrical conduction by those thermally inactive bands is ignored in the present calculation.
In this framework, we express $\gzzzz$ as the sum of orbital ($g_{zzzz}^\mathrm{orb}$) and spin ($g_{zzzz}^\mathrm{spin}$) parts, and $g_{zzzz}^\mathrm{orb}$ is given by~\cite{SM}
\begin{align}
    \gzzzz^\mathrm{orb} &= 
    \int\frac{\mathrm{d}\kk}{(2\pi)^3}
    \mathfrak{g}_{zzzz}^\mathrm{orb}(\kk)
    \left.\left(-\frac{\partial f_0}{\partial E}\right)\right|_{E=E_\mathrm{v}},
\end{align}
where $\mathfrak{g}_{zzzz}^\mathrm{orb}(\kk) =-(2e^3/5\hbar^2)\tau_\kk^2 \partial_{k_z}[v_z^2\partial_{k_z}(m_z^\mathrm{orb}/v_z)]$.
Here, 
$v_z = (1/\hbar) \partial_{k_z} E_\mathrm{v}$ and $f_0(E)$ denotes the Fermi-Dirac distribution function for electrons. $\tau_\kk$ represents an effective relaxation time for NEC~\cite{SM}. 
$g_{zzzz}^\mathrm{spin}$ is determined in the same way. 
We focus on the absolute sign by considering $\mathfrak{g}_{zzzz}^\mathrm{orb,spin}(\kk)$ in the $k_z$--$k_x$ plane for left-handed Te [Figs.~\ref{fig:g_frak} (a), (b)]. 
$\mathfrak{g}_{zzzz}^\mathrm{orb}$ is positive while $\mathfrak{g}_{zzzz}^\mathrm{spin}$ is negative, originating from the opposite signs of $m_z^\mathrm{orb}$ and $m_z^\mathrm{spin}$.
Notably, the sum $\mathfrak{g}_{zzzz}=\mathfrak{g}_{zzzz}^\mathrm{orb}+\mathfrak{g}_{zzzz}^\mathrm{spin}$ is positive over the entire $\bm{k}$ space around the H point [Fig.~\ref{fig:g_frak}(c)], resulting in a positive sign of $g_{zzzz}$ within the accessible energy range of our measurement: the same sign as left-handed $\Gzzzz>0$. 
Because the orbital and spin textures change sign with the handedness of Te, the present argument holds equally for right-handed $\Gzzzz<0$.
The agreement in enantiospecific sign demonstrates that NEC primarily originates from the orbital texture. 
Contributions of the Berry curvature ($\mathfrak{g}_{zzzz}^\mathrm{BC}$) were not considered in the calculations. 
Boltzmann kinetic theory combined with {\it ab initio} calculations~\cite{liu2023electrical} showed that $\mathfrak{g}_{zzzz}^\mathrm{BC}$ is even smaller than $\mathfrak{g}_{zzzz}^\mathrm{spin}$, because $\mathfrak{g}_{zzzz}^\mathrm{BC}$ is directly proportional to the low group velocity, unlike $\mathfrak{g}_{zzzz}$, alongside the absence of topological singularities.

{\it Carrier density dependence of $G_{zzzz}$.}---
To investigate the carrier-density dependence of $\Gzzzz$,
we fabricated a Te crystal into a field-effect transistor [device C in Fig.~\ref{fig:carrier_dep}(a)], where longitudinal and Hall voltages were measured to calculate $G_{zzzz}$ and the sheet carrier density $n_\mathrm{S}$ at various bottom-gate voltages. 
By varying the magnetic field angle $\alpha$ in the $yz$ plane, we were able to precisely determine $\Gzzzz$ and $n_\mathrm{S}$ by fitting with $\cos\alpha$ and $\sin\alpha$, respectively, as illustrated in Fig.~\ref{fig:carrier_dep}(b).

The dependence of $G_{zzzz}$ on $n_\mathrm{S}$ is shown in Fig.~\ref{fig:carrier_dep}(c). $G_{zzzz}$ decreases with decreasing $n_\mathrm{S}$ at a temperature $T$ of 20 K, which is the lowest temperature of our measurement system. 
However, the magnitude of the slope decreases below $n_\mathrm{S}\sim17\times10^{12}~\mathrm{cm^{-2}}$ and increases again below $n_\mathrm{S}\sim8\times10^{12}~\mathrm{cm^{-2}}$. Consequently, $\Gzzzz$ exhibits a broad peak structure in the low carrier density range. This peak structure weakens and disappears with increasing $T$. Figure~\ref{fig:carrier_dep}(d) demonstrates that these observed trends are consistent with calculated $\gzzzz$, where the chemical potential and temperature were considered via $f_0(E)$. 
Division of $\gzzzz$ into $\gzzzz^\mathrm{orb}$ and $\gzzzz^\mathrm{spin}$ reveals that the peak structure is primarily caused by $\gzzzz^\mathrm{orb}$ rather than $\gzzzz^\mathrm{spin}$ in the carrier-density dependence [see the inset to Fig.~\ref{fig:carrier_dep}(d)]. 
Therefore, our findings suggest that $\Gzzzz$ reflects the energy dependence of $m_z^\mathrm{orb}$ embedded in $\gzzzz^\mathrm{orb}$, which is concentrated around the valence band top (see also Fig.~\ref{fig:chiral_properties}).
Within the present temperature range, charge carriers experience non-negligible scattering due to electron-phonon and electron-electron interactions, not accounted for in our calculations based on elastic impurity scattering. 
Consequently, the calculated carrier-density dependence of NEC would be less apparent in experiments, possibly due to Matthiessen's rule of the relaxation times. 

Our calculations did not consider the extrinsic orbital magnetic moment $\bm{m}^{\mathrm{orb,ext}}$, arising from the antisymmetric impurity scattering~\cite{csahin2018pancharatnam}.
Because scattering processes responsible for $m_z^{\mathrm{orb,ext}}=\bm{m}^{\mathrm{orb,ext}}~\cdot \bm{e}_z$ are enhanced with increasing $k_{x,y}$, $m_z^{\mathrm{orb,ext}}$ is enhanced with increasing carrier density, or expanding Fermi surface. 
Theoretically, the current-induced orbital magnetization of Te changes from intrinsic to extrinsic with increasing carrier density, and a similar carrier-density dependence to Fig.~\ref{fig:carrier_dep}(c) is exhibited by the conversion efficiency~\cite{csahin2018pancharatnam}. 
The crossover may be relevant to our results, explaining the monotonic increase of $\Gzzzz$ for higher $n_\mathrm{S}$.
Even for this mechanism, there exists a low carrier density range in which $k_{x,y}$ is much smaller than $k_z$, and NEC discernibly derives from the orbital magnetic moment determined by the band structure. 
We interpret the observed broad peak structure as an indication of such a carrier density range.

\begin{figure}[ht]
\begin{center}
\includegraphics[width=85mm]{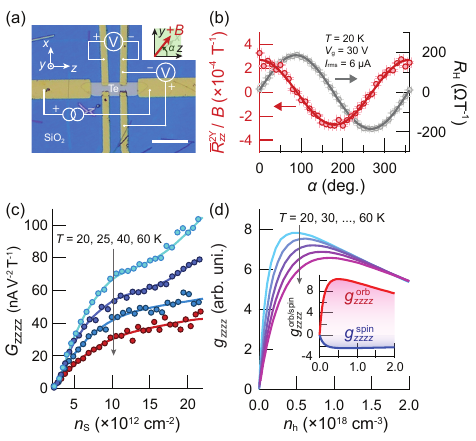}
\end{center}
\caption{
(a) Microscope image of device C. Scale bar: 30~\textmu m.
(b) Dependences of $\RY$ and the Hall coefficient $R_\mathrm{H}$ on $\alpha$, measured at back gate voltage $V_\mathrm{g}=30$~V and temperature $T=20$~K.
(c) Nonlinear conductance $\Gzzzz$ as a function of the sheet carrier density $n_\mathrm{S}$. Lines are guides to the eye.
(d) Nonlinear conductivity $\gzzzz$ as a function of the hole carrier density $n_\mathrm{h}$. The orbital component $\gzzzz^\mathrm{orb}$ and the spin component $\gzzzz^\mathrm{spin}$ at 20~K are also shown in the inset.
}
\label{fig:carrier_dep}
\end{figure}

{\it Summary.}---
We investigated the nonlinear electrical conduction (NEC) in chiral semiconductor Te, and uncovered that NEC primarily stems from the chiral orbital texture of a Bloch electron. 
This was evidenced by the enantiospecific sign of the nonlinear electrical conductance and anomalies in its carrier-density dependence. These experimental characteristics were rationalized by integrating semiclassical Boltzmann kinetic theory, accounting for the orbital texture via the Zeeman interaction. Our findings underscore the crucial role of orbital effects in inducing NEC and highlight the potential of NEC measurements in probing orbital textures across various materials.

{\it Note added.}---
Immediately before submitting the original version of our manuscript, we became aware of the theoretical work by K. Nakazawa {\it et al.} based on {\it ab initio} calculations~\cite{nakazawa2024nonlinear}, who calculated another type of nonlinear charge transport driven by the product of the electric field and the temperature gradient. They identified the orbital magnetic moment as the primary origin of this effect in the vicinity of the valence band top.

\acknowledgements{
We thank Tomohiro Seki for technical assistance in sulfuric acid etching. 
We are also grateful to Masashige Matsumoto for his discussions.
We acknowledge Tetsu Mieno for the support of scanning electron microscopy provided by Molecular Structure Analysis Section, Shizuoka Instrumental Analysis Center, Shizuoka University.
This work was supported by 
Grant-in-Aid for Scientific Research (B) (23H01836),
for Challenging Research (Exploratory) (22K18695, 24K21527),
and for Transformative Research Areas (A) (24H02234)
from JSPS KAKENHI, Japan as well as 
PRESTO "Topological Materials Science for Creation of Innovative Functions" (JPMJPR20L9) from JST, Japan.
}


%

\onecolumngrid
\newpage
\begin{center}\large
   \textbf{Supplemental Material for\\“Chiral orbital texture in nonlinear electrical conduction"}
\end{center}

\begin{center}
    Suguru Okumura, Ryutaro Tanaka, and Daichi Hirobe
    
    \textit{Shizuoka University, Suruga, Shizuoka, 422-8529, Japan.}
\end{center}
\maketitle

\subsection{Parameters used in the tight-binding model}
Following Ref.~48, we present the parameters of the tight-binding model in which the effective Hamiltonian acts on the reduced space spanned by two conduction bands and four valence bands.
The conduction band $E_\mathrm{c}$, the highest valence band $E_\mathrm{v}$, and the lowest valence band $E_\mathrm{h}$ are parameterized as
\begin{align}
    E_\mathrm{c}(\kk) &= 
    + \frac{\hbar^2 (k_x^2 + k_y^2)}{2m_\perp^\mathrm{c}} 
    + \frac{\hbar^2 k_z^2}{2m_{\parallel}^\mathrm{c}} 
    + E_\mathrm{g},\\
    E_\mathrm{v}(\kk) &= 
    - \frac{\hbar^2 (k_x^2 + k_y^2)}{2m_\perp^\mathrm{v}} 
    - \frac{\hbar^2 k_z^2}{2m_{\parallel}^\mathrm{v}} 
    + \sqrt{(Sk_z)^2 + \varDelta^2}-\varDelta,\\
    E_\mathrm{h}(\kk) &= 
    - \frac{\hbar^2 (k_x^2 + k_y^2)}{2m_\perp^\mathrm{h}} 
    - \frac{\hbar^2 k_z^2}{2m_{\parallel}^\mathrm{h}} 
    -2\varDelta' - \varDelta.
\end{align}
We omitted the expression for the second highest valence band, which does not contribute to the orbital magnetic moment for the highest valence band due to the symmetry constraint.
The parameter values are listed in Table SI.

The spin magnetic moment $m_z^\mathrm{spin}$ and the orbital magnetic moment $m_z^\mathrm{orb}$ for the highest valence band are given by
\begin{align}
    m_z^\mathrm{spin} 
    &= -\mu_\mathrm{B}\frac{Sk_z}{\sqrt{(Sk_z)^2+\varDelta^2}}
    = -\mu_\mathrm{B}\eta(k_z), \\
    m_z^\mathrm{orb}
    &= -\frac{\mu_B}{2\hbar c \lambdabar_\mathrm{C}}
    \left(
        \frac{|P_\mathrm{vc}^+|^2-|P_\mathrm{vc}^-|^2}{E_\mathrm{v}-E_\mathrm{c}}
        + \frac{|P_\mathrm{vh}^+|^2-|P_\mathrm{vh}^-|^2}{E_\mathrm{v}-E_\mathrm{c}}
    \right)
    \eta(k_z).
\end{align}
$c$ is the speed of light, $\lambdabar_\mathrm{C}$ is the reduced Compton wavelength of the electron in vacuum, and $P_\mathrm{vc(vh)}^\pm = \bra{u_\mathrm{v}}(\partial_{k_x} \pm i \partial_{k_y})\ket{u_\mathrm{c(h)}}$ with $\ket{u_\mathrm{v,c,h}}$ being the corresponding energy eigenstates in the reduced space.
By setting $\epsilon_\mathrm{c(h)}=(|P_\mathrm{vc(vh)}^+|^2-|P_\mathrm{vc(vh)}^-|^2)/(2\hbar c \lambdabar_\mathrm{C})$, we adjusted $\epsilon_\mathrm{c(h)}$ to reproduce the $k_z$ dependence of the orbital magnetic moment obtained by {\it ab initio} calculations~[47]. 
As shown in Fig.~S1,
good agreement is obtained for $\epsilon_\mathrm{c}=4.965$ eV and $\epsilon_\mathrm{h}=1.745$ eV, and the magnitudes are consistent with the original ones~[48].
We note that the enantiomeric sign of $m_z^\mathrm{spin}$ agrees with that determined by spin- and angle-resolved photoelectron spectroscopy~[21] as well as nuclear magnetic resonance spectroscopy~[30].

\renewcommand{\thetable}{SI}
\begin{table}[h]
    \begin{tabular}{ c||l|c||l }
        \hline
        $\hbar^2/(2m_\perp^\mathrm{c})$     & 57.5~eV\AA$^2$  & $\varDelta$                         & 0.063~eV \\
        $\hbar^2/(2m_\parallel^\mathrm{c})$ & 52.9~eV\AA$^2$  & $\hbar^2/(2m_\perp^\mathrm{h})$     & 46.0~eV\AA$^2$ \\
        $E_\mathrm{g}$                      & 0.335~eV        & $\hbar^2/(2m_\parallel^\mathrm{h})$ & 34.3~eV\AA$^2$ \\
        $\hbar^2/(2m_\perp^\mathrm{v})$     & 32.6~eV\AA$^2$  & $\varDelta'$                        & 0.053~eV \\
        $\hbar^2/(2m_\parallel^\mathrm{v})$ & 36.4~eV\AA$^2$  & $\epsilon_\mathrm{c}$               & 4.965~eV \\
        $S$                                 & $\pm2.30$~eV\AA & $\epsilon_\mathrm{h}$               & 1.745~eV \\
        \hline
    \end{tabular}
    \caption{Parameters of the energy dispersions and the magnetic moments. The sign of $S$ is positive for right-handed Te ($P3_121$) and negative for left-handed Te ($P3_221$) in accordance with {\it ab initio} calculations~[21, 47] as well as experimental results for spin- and angle-resolved photoelectron spectroscopy~[21] and nuclear magnetic resonance spectroscopy~[30].}
\end{table}

\renewcommand{\thefigure}{S1}
\begin{figure}
\centering
        \includegraphics[width=110mm]{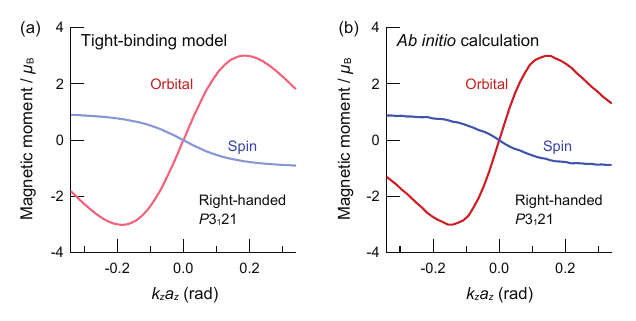}
\caption{ 
Quasimomentum ($k_z$) dependences of orbital ($m_z^\mathrm{orb}$) and spin ($m_z^\mathrm{spin}$) magnetic moments in the framework of the tight-binding model (a)~[48] and {\it ab initio} calculations (b)~[47]. $a_z$ is the lattice constant along the $c$ axis of Te.
}
\end{figure}

\newpage

\subsection{Hydrothermal synthesis and lamination of Te crystals}
Na$_2$TeO$_3$ (53.4 mg) was dissolved in ultrapure water (17.6 mL) at room temperature under magnetic stirring to form a homogeneous solution. 
Subsequently, polyvinylpyrrolidone (272.6 mg) and an aqueous ammonia solution 25\% (1 mL) were added to the solution, and the mixture was stirred thoroughly with the magnetic stirrer. 
3 mL of the mixture was transferred to a 4mL autoclave, and hydrazine monohydrate 98\% (0.1 mL) was added to the mixture. 
The autoclaves were sealed, maintained at a reaction temperature of 180 $^{\circ}$C for 28h to synthesize Te crystals, and cooled down naturally to room temperature. 
This reaction process was repeated twice to produce thick Te crystals.
Te crystals were precipitated by centrifugation at 5000 rpm for 10 minutes, followed by rinsing with ultrapure water three times to remove any ions remaining in the final products.
The Te crystals were redispersed in ethanol, drop-casted onto the SiO$_2$ gate dielectric (300 nm thick), and dried naturally to complete the lamination, followed by solvent cleaning with acetone and isopropanol. Prior to drop-casting, the Si substrates with the gate dielectric were rinsed with acetone and isopropanol, followed by cleaning through ozone irradiation.

\subsection{Etching by sulfuric acid and scanning electron microscope images of etch pits over wide area}
Slow etching by sulfuric acid was conducted with modifications to the method in Ref.~31: specifically, we immersed samples in sulfuric acid on the hotplate at a set temperature of 100 $^{\circ}$C and decreased the immersion time to 30 seconds from 1,800 seconds in Ref.~31. With the modifications, we were able to restrict etching to the surface of tellurium, which was confirmed by the spatial mapping of the etch pits by scanning electron microscopy~(Fig.~S2). We also note that the spatially uniform shape of the asymmetric etch pits demonstrates the enantiopure nature of the crystal used.

\renewcommand{\thefigure}{S2}
\begin{figure}[ht]
\centering
        \includegraphics[width=110mm]{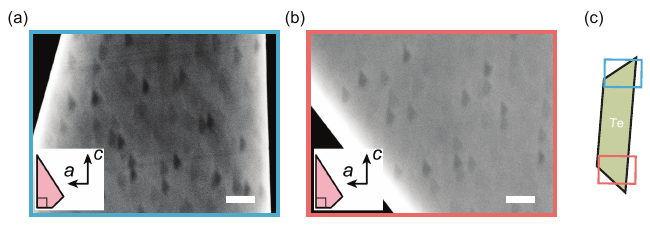}
\caption{
(a), (b) Scanning electron microscope images of asymmetric etch pits of device B. Scale bar: 1~\textmu m.
(c) Schematic of the areas of the scanning electron microscope images taken.
}
\end{figure}

\subsection{Temperature-variable harmonic resistances of devices A and B}
In Fig.~S3, we show the temperature dependence of the harmonic resistance $R^\mathrm{1X}_{zz}$ for devices A and B. Semiconducting and metallic behaviors are exhibited by devices A and B, respectively, whose difference could be attributed to unintentional hole doping in the crystal growth of tellurium. 
The result shows that the chemical potential is higher for device A, allowing for more  selective thermal excitation near the valence band maximum of tellurium. Therefore, device A is expected to exhibit the higher rectification efficiency [18, 33], which is consistent with the higher value of $\RY/B$ for device A in Fig. 2 in the main text.

\renewcommand{\thefigure}{S3}
\begin{figure}[ht]
\centering
        \includegraphics[width=65mm]{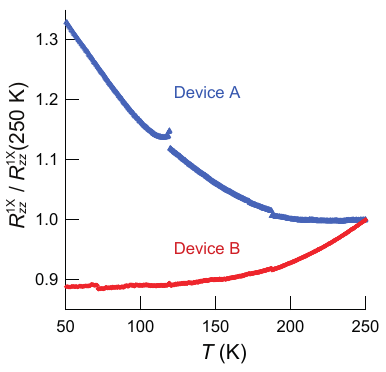}
\caption{
Temperature $T$ dependence of the harmonic longitudinal resistance $R^\mathrm{1X}_{zz}$ for devices A and B.
Values of the resistances are normalized by those values at 250 K (41~k$\Omega$ for device A and 8.3~k$\Omega$ for device B at 250 K).
}
\end{figure}

\subsection{Comparison between four- and two-terminal measurements of nonlinear electrical conduction}
To compare between four- and two-terminal measurements of nonlinear electrical conduction, we prepared a four-terminal device (device D, whose handedness was not determined), and measured $\RY/B$ as a function of the magnetic field angle $\beta$ at 30 K in the same manner as Fig. 2(e) in the main text. It may be noted that the expression for $\RY/B$ does not contain the channel length, which allows for direct comparison between four- and two-terminal configurations.
By fitting to the datasets in Fig.~S4 with $\cos\beta$, the signed amplitude is found to be $-(1.53\pm0.11)\times10^{-4}$~T$^{-1}$ for four-terminal configuration and $-(1.52\pm0.05)\times10^{-4}$~T$^{-1}$ for two-terminal configuration. The excellent agreement shows the negligibly small contribution of extrinsic nonlinear transport due to contact resistance. Therefore, one can apply two-terminal measurement to discuss the absolute enantiomeric sign of the nonlinear conductance of tellurium.

\renewcommand{\thefigure}{S4}
\begin{figure}[ht]
\centering
        \includegraphics[width=130mm]{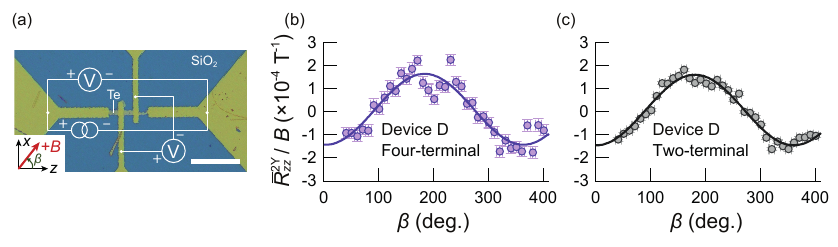}
\caption{
(a) Microscope image of device D. Scale bar: 50~\textmu m. 
(b), (c) Magnetic field angle $\beta$ dependence of the normalized second harmonic resistance $\RY$ for four-terminal (b) and two-terminal (c) configurations at 30 K. The excitation electric current $I_\mathrm{rms}$ was set to 10~\textmu A.
}
\end{figure}

\subsection{Rewriting of $\RY$ in terms of $\Gzz$ and $\Gzzzz$}
We start with the phenomenological relation
\begin{equation}
    j_z = \gzz E_z + \gzzzz (E_z)^2 B_z.
\end{equation}
It follows that
\begin{align}
    E_z 
    &= (\gzz)^{-1}j_z - (\gzz)^{-1}\gzzzz (E_z)^2 B_z \notag \\
    &= (\gzz)^{-1}j_z - (\gzz)^{-1}\gzzzz \{(\gzz)^{-1}j_z - (\gzz)^{-1}\gzzzz (E_z)^2 B_z\}^2 B_z \notag \\
    &= (\gzz)^{-1}j_z - (\gzz)^{-3}\gzzzz (j_z)^2B_z + \Order((j_z)^3).
\end{align}
For the experimental setup shown in the main text, $E_z$ generates a voltage drop $V_z$ over a length $l_z$ along the $z$ axis:
\begin{equation}
    V_z = \int_0^{-l_z}\mathrm{d}z(-E_z)=E_zl_z= V_z^{(1)} + V_z^{(2)}.
\end{equation}
Superscripts (1) and (2) denote the order with respect to $j_z$. Using $j_z = I_z / S_z$ and $\gzz = \Gzz l_z/S_z$ ($S_z$: the cross-section of the sample perpendicular to the $z$ axis), we obtain
\begin{equation}
    V_z^{(2)} 
    = -\frac{S_z}{(l_z)^2} (\Gzz)^{-3} \gzzzz (I_z)^2B_z
    = - (\Gzz)^{-3} \Gzzzz (I_z)^2B_z.
\end{equation}
where $\Gzzzz = \gzzzz S_z/(l_z)^2$.

We rewrite $V_z^{(2)}$ in order to compare with experimental voltage signals in phase-sensitive detection. 
Subsequently, we use the root mean square (rms) for a.c. signals.
For phase-sensitive detection, a sinusoidal electric current $I_z(t)=\sqrt{2}I_{\mathrm{rms}}\sin(\omega t)$ with $\omega = 2\pi f$ is sourced to a device under test.
The substitution of $I_z(t)$ in $V_z^{(2)}$ yields
\begin{align}
    V_z^{(2)}(t) 
    &= (\Gzz)^{-3} \Gzzzz (I_\mathrm{rms})^2 
    \left\{
        -1+\sin\left(2\omega t+\frac{\pi}{2}
    \right)\right\}B_z.
\end{align}
Therefore, a second-harmonic voltage, $\VY$, in rms with a phase shift of $+\pi/2$ (i.e. out of phase) is given by
\begin{equation}
    \VY = \frac{1}{\sqrt{2}} (\Gzz)^{-3} \Gzzzz (I_\mathrm{rms})^2 B_z.
\end{equation}
Because $\VX = (\Gzz)^{-1}I_\mathrm{rms}$, we finally have 
\begin{equation}
    \RY 
    = \frac{\VY}{\VX} 
    = \frac{1}{\sqrt{2}}(\Gzz)^{-2}\Gzzzz I_\mathrm{rms}B_z.
\end{equation}

We comment on the calculation of $\RY$ on account of pulse amplitude modulation 
of excitation electric current in phase-sensitive detection. 
When the duty cycle of pulse amplitude modulation is set to $D$ ($0<D<1$), 
experimentally detected $\VX$ and $\VY$ are both multiplied by $D$. Therefore, $\RY=\VY/\VX$ is independent of $D$. 

\subsection{Double-checking of the sign of $\Gzzzz$ by d.c. magnetoconductance measurement}
The sign of $\Gzzzz$ determined by phase-sensitive detection was double-checked by d.c. measurement. Magnetoconductance measurements were performed by applying a d.c. voltage $V_z$ to device B (left-handed, $P3_221$) along the $z$-axis and measuring a d.c. electric current $I_z$ in the same direction. A magnetic field $B_z$ was also applied along the $z$ axis. The magnitude of $V_z^{\mathrm{dc}}$ was set to $0.25$~V. A source measure unit (Keithley 2636B) was used for the electric d.c. excitation and measurement. The d.c. electrical conductance $G$ was calculated as $G(V_z) = I_z / V_z$. 

In Fig.~S5, we show a conductance difference $\Delta G = [G(+|V_z|)-G(-|V_z|)]/2$ as a function of $B_z$. A conductance sum $\overline{G} = [G(+|V_z|)+G(-|V_z|)]/2$ is also shown for comparison.  $\Delta G$ is proportional to $B_z$, and the sign of $\Delta G$ is reversed upon reversal of $B_z$. This $B_z$-regulated unipolarity of $\Delta G$ is a characteristic of nonlinear electrical conduction. Because $G$ is approximated by $G(V_z) = \Gzz + \Gzzzz V_zB_z$ in the low magnetic field range, we have $\Delta G = \Gzzzz |V_z| B_z$. Therefore, the positive slope of $\Delta G$ vs $B_z$ indicates the positive sign of $\Gzzzz$ for device B (left-handed, $P3_221$), which agrees with the result obtained by phase-sensitive detection. The agreement validates the analytical procedures in the preceding section.

\renewcommand{\thefigure}{S5}
\begin{figure}[ht]
\centering
        \includegraphics[width=110mm]{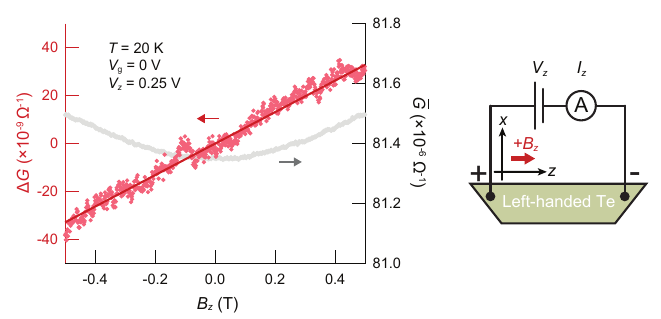}
\caption{ 
Magnetic field $B_z$ dependences of a d.c. conductance difference $\Delta G$ and a d.c. conductance sum $\overline{G}$, measured at temperature $T=20$~K and back gate voltage $V_\mathrm{g}=0$~V. The magnitude of the source voltage $V_z$ was set to 0.25~V.
}
\end{figure}

\subsection{Calculation of $\gzzzz$ by Boltzmann kinetic equation}
\subsubsection{Boltzmann kinetic equation for an electron wavepacket}
We use semiclassical Boltzmann kinetic theory to derive a nonequilibrium distribution function in quasimomentum $(\kk)$ space under electric and magnetic fields, and calculate $\gzzzz^\mathrm{spin}$ for the valence band of Te. We will see that $\gzzzz^\mathrm{orb}$ is derived in a similar manner. We caution that the dynamics of electrons, not holes, is treated in the following calculations to prevent possible confusion about the sign of the nonlinear conductivities. 

We start with the following equation of motion for an electron wavepacket:
\begin{align}
    \dot{\bm{r}} &= \frac{1}{\hbar}\nabla_\kk \widetilde{E}_\mathrm{v}-\dot{\kk}\times\bm{\varOmega}(\kk), \\
    \hbar\dot{\kk} &= -e(\bm{E}+\dot{\bm{r}}\times\bm{B}),
\end{align}
where $\widetilde{E}_\mathrm{v}=E_\mathrm{v}-\bm{m}^\mathrm{spin} \cdot \bm{B}$ denotes the valence band plus spin Zeeman interaction, and $\bm{\varOmega}(\kk)$ the Berry curvature. The elementary charge $e=|e|$ is positive in our notation. Decoupling $\dot{\bm{r}}$ and $\dot{\bm{k}}$ yields
\begin{align}
    D\dot{\bm{r}} &= \frac{1}{\hbar}\left[
        \nabla_\kk \widetilde{E}_\mathrm{v} + 
        e\bm{E}\times\bm{\varOmega}(\kk) + \frac{e}{\hbar}\left(\nabla_\kk \widetilde{E}_\mathrm{v}\cdot\bm{\varOmega}(\kk)\right)\bm{B}
    \right], \\
    D\dot{\kk} &= -\frac{e}{\hbar}\left[
        \bm{E}
        + \frac{1}{\hbar}\nabla_\kk \widetilde{E}_\mathrm{v} \times \bm{B}
        + \frac{e}{\hbar} \left(\bm{E}\cdot\bm{B}\right)\bm{\varOmega}(\kk)
    \right].
\end{align}
Here $D = 1 + (e/\hbar)\bm{B}\cdot\bm{\varOmega}(\kk)$.
For the spatially uniform system, an electric current density $\bm{j}$ is given by
\begin{equation}
    \bm{j} = (-e)\int\frac{\mathrm{d}\kk}{(2\pi)^3}D\dot{\bm{r}} f,
\end{equation}
where $f$ denotes a nonequilibrium distribution function for electrons in $\kk$ space.

We next calculate $f$ for a steady state which satisfies the Boltzmann kinetic equation
\begin{equation}
    \dot{\kk}\cdot \dkf + \Integ^{(\mathrm{el})}[f]=0.
\end{equation}
$\Integ^{(\mathrm{el})}[f]$ denotes the collision integral of elastic scattering. Because the scattering rate of momentum relaxation is much larger than that of energy relaxation at sufficiently low temperatures, we will ignore the collision integral of inelastic scattering when discussing the sign of low-temperature nonlinear electrical conductivity. 

\subsubsection{Collision integral in the relaxation-time approximation}
We temporarily use the constant--$\tau$ approximation for $\Integ^\mathrm{(el)}[f]$. However, this simplified model could be problematic for band edges and/or anisotropic energy bands~\cite{schliemann2003anisotropic, vyborny2009semiclassical, liu2016mobility, xiao2016unconventional, kim2019vertex}, both of which apply to the dumbbell structure near the valence band top of Te.
Therefore, the present section should be viewed for argument's sake only.

Taking $\Integ^\mathrm{(el)}[f]=(f-f_0)/\tau_0$ with $\tau_0$ being constant, we have
\begin{equation}
    \dot{\kk}\cdot\nabla_\kk f = -\frac{f-f_0}{\tau_0},
\end{equation}
where $f_0$ is the equilibrium distribution function for electrons.
The second-order term of $E_z$ in $f$, $f_2$, is responsible for the nonlinear electric current. 
The successive substitution of $f$ in the Boltzmann equation leads to 
\begin{equation}
    f_2 
    = 
    (\tau_0)^2(\dot{\kk}\cdot\nabla_\kk)^2f_0
    = 
    (\tau_0)^2\left(\frac{e}{\hbar D}\right)^2
    \left\{
        \left[
            \bm{E}+\frac{e}{\hbar}(\bm{E}\cdot\bm{B})\bm{\varOmega}(\kk)
        \right]\cdot\nabla_\kk
    \right\}^2f_0.
\end{equation}
The same approach was taken in Ref.~18 in combination with first-principles calculations. The result showed that the Berry curvature plays a minor role in the nonlinear electric current for the valence band of Te in our experimental geometry. Therefore, we will ignore terms which contain $\bm{\varOmega}(\kk)$ in the following calculations. The nonlinear electric current $\bm{j}^{(2)}$ is approximated by 
\begin{align}
    \bm{j}^{(2)}
    &= (-e)\int\frac{\mathrm{d}\kk}{(2\pi)^3}D\dot{\bm{r}} f_2 \notag \\
    &\approx (-e)\left(\frac{\tau_0 e}{\hbar}\right)^2 \int \frac{\mathrm{d}\kk}{(2\pi)^3}
    \dot{\bm{r}} (\bm{E}\cdot\nabla_\kk)^2 f_0 \notag \\
    &=+\frac{(\tau_0)^2e^3}{\hbar^3}\int\frac{\mathrm{d}\kk}{(2\pi)^3}
    \left[(\bm{E}\cdot\nabla_\kk)\nabla_\kk \widetilde{E}_\mathrm{v}\right]
    (\bm{E}\cdot\nabla_\kk \widetilde{E}_\mathrm{v})
    \frac{\partial f_0}{\partial \widetilde{E}_\mathrm{v}}.
\end{align}
Partial integration was performed in the last line. 
For its relevance to our experiment, a component of $\bm{j}^{(2)}$, $\bm{j}^{(2)}_{E^2B}$, will be considered which is of the second order of $\bm{E}$ and the first order of $\bm{B}$.
Direct computation shows
\begin{equation}
    \bm{j}^{(2)}_{E^2B} 
    =\frac{(\tau_0)^2e^3}{\hbar^2}\int\frac{\mathrm{d}\kk}{(2\pi)^3}
    \left\{
    (\bm{E}\cdot\bm{v})
    \left[(\bm{E}\cdot\nabla_\kk)\nabla_\kk(\bm{m}^\mathrm{spin}\cdot\bm{B})\right]
    -\left[(\bm{E}\cdot\nabla_\kk)^2\bm{v}\right](\bm{m}^\mathrm{spin}\cdot\bm{B})
    \right\}
    \left.\left(-\frac{\partial f_0}{\partial E}\right)\right|_{E=E_\mathrm{v}}.
\end{equation}
Here $\bm{v}=(1/\hbar)\nabla_\kk E_\mathrm{v}$ denotes the group velocity for the valence band without Zeeman interaction. When $\bm{E}=E_z\bm{e}_z$ and $\bm{B}=B_z\bm{e}_z$, 
we find $\bm{j}^{(2)}_{E^2B} \cdot \bm{e}_z = \gzzzz^\mathrm{spin}(E_z)^2B_z$. $\gzzzz^\mathrm{spin}$ is given by
\begin{align}
    \gzzzz^\mathrm{spin} &= 
    \int\frac{\mathrm{d}\kk}{(2\pi)^3}
    \mathfrak{g}_{zzzz}^\mathrm{spin}(\kk)
    \left.\left(-\frac{\partial f_0}{\partial E}\right)\right|_{E=E_\mathrm{v}}, \\
    \mathfrak{g}_{zzzz}^\mathrm{spin}(\kk) 
    &= (\tau_0)^2\frac{e^3}{\hbar^2}
    \frac{\partial}{\partial k_z}
    \left[
        (v_z)^2\frac{\partial}{\partial k_z}
        \frac{m_z^\mathrm{spin}}{v_z}
    \right].
\end{align}
The orbital counterparts are obtained by replacing $m_z^\mathrm{spin}$ with $m_z^\mathrm{orb}$.  

\subsubsection{Collision integral beyond the relaxation-time approximation}
In Ref.~57, the authors developed the formalism of nonlinear electrical conduction based on the inverse collision integral on account of the anisotropic valence band of Te. 
We outline key results of the formalism beyond the constant--$\tau$ approximation and revise the expression for the nonlinear electrical conductivity introduced in the previous section. 

One considers an electric current which is subject to elastic scattering by short-range impurity potential $U(\rr)=U_0\delta(\rr)$. $\Integ^{(\mathrm{el})}[f]$ is approximated by
\begin{align}
    \Integ^{(\mathrm{el})}[f] &= \frac{2\pi}{\hbar}n_\mathrm{i}
    \int\frac{\mathrm{d}\kk'}{(2\pi)^3}\delta\left(\Ev(\kk)-\Ev(\kk')\right)|\bra{\kk}U\ket{\kk'}|^2\{f(\kk)-f(\kk')\},\\
    \bra{\kk}U\ket{\kk'} &= e^{i(\kk'-\kk)\cdot\rr}U_0\braket{u_\mathrm{v}(\kk)|u_\mathrm{v}(\kk')}.
\end{align}
$n_i$ is a dilute impurity density. When external magnetic and electric fields are absent, the wavefunction $u_\mathrm{v}(\kk)$ for $E_\mathrm{v}$ reads $u_\mathrm{v}(\kk)=[\sqrt{(1+\eta)/2}, \sqrt{(1-\eta)/2}]^\mathrm{t}$. It follows that
\begin{equation}
    |\bra{\kk}U\ket{\kk'}|^2 = \frac{U_0^2}{2}\{1+\eta(k_z)\eta(k'_z)+\zeta(k_z)\zeta(k'_z)\},
\end{equation}
where $\zeta = \sqrt{1-\eta^2}$.
In Ref.~57, the authors directly constructed the inverse operator of the collision integral without presupposing beforehand the relaxation time. They showed that an effective relaxation time can be derived from this operator as the leading-order term in the limit that $|B k_z/\Delta| \ll 1$, which is valid for $\kk$ space near the valence band top of Te. By calculating corrections of $\Integ^{(\mathrm{el})}[f]$ due to magnetic and electric fields and combining with the group velocity modified by Zeeman interaction, the authors essentially showed that $\gzzzz^\mathrm{spin}(\kk)$ is given by
\begin{align}
    \gzzzz^\mathrm{spin} &= 
    \int\frac{\mathrm{d}\kk}{(2\pi)^3}
    \mathfrak{g}_{zzzz}^\mathrm{spin}(\kk)
    \left.\left(-\frac{\partial f_0}{\partial E}\right)\right|_{E=E_\mathrm{v}}, \\
    \mathfrak{g}_{zzzz}^\mathrm{spin}(\kk) 
    &=
    -\frac{2}{5}
    \left(\tau_\kk^\mathrm{F}\right)^2
    \frac{e^3}{\hbar^2}
    \frac{\partial}{\partial k_z}
    \left[
        (v_z)^2\frac{\partial}{\partial k_z}
        \frac{m_z^\mathrm{spin}}{v_z}
    \right]
\end{align}
in the absolute zero temperature limit.
$\tau_\kk^\mathrm{F}$ is an effective relaxation time at the Fermi level defined as $\tau_\kk^\mathrm{F} \approx 2 \pi (\hbar^2/2m_\perp^\mathrm{v}) \hbar/[n_\mathrm{i}U_0^2 \kappa_z(E_\mathrm{F})]$
with $\kappa_z(E)$ being the maximum value of $k_z$ at $E_\mathrm{v}=E$.
The energy dependence of $\tau_\kk^\mathrm{F}$ via $\kappa_z$ reflects the fact that the energy dispersion is so elongated along the $k_z$ axis that electrical resistance is generated dominantly by backscattering parallel to the $k_z$ axis, which carries by far the largest quasimomentum change [see also the Fermi surface in Fig.~1(c) of the main text].

Subsequently, we explain approximations made for our model calculations.
As noted in Ref.~57, the analytical expression for $\tau_\kk^\mathrm{F}$ was derived under the assumption that the energy scale of charge carriers is larger than the energy maximum $E_\mathrm{v}^\mathrm{max}$, namely the height of the dumbbell structure. 
Therefore, one should consider that $\tau_\kk^\mathrm{F}$ is asymptotic to $\tau_\kk^\mathrm{F} \propto 1/\kappa_z(E_\mathrm{F})$ as the Fermi level moves downward from the dumbbell structure.
For the extrapolation of $\tau_\kk^\mathrm{F}$ to the dumbbell structure, $\tau_\kk^\mathrm{F}$ is assumed to converge to a constant $\tau_0$ as $E_\mathrm{F}$ approaches $E_\mathrm{v}^\mathrm{max}$.
The assumption allows us to set $\tau_\kk^\mathrm{F}=\tau_0\tanh(\kappa_z^0/\kappa_z^\mathrm{F})$ with $\kappa_z^0$ being on the order of $\kappa_z(E=0)$.
As long as the asymptotic behavior of $\tau_\kk^\mathrm{F}$ is maintained, other approximate forms of $\tau_\kk^\mathrm{F}$ will equally suffice.

Although the original calculations were performed for $\mathfrak{g}_{zzzz}^\mathrm{spin}(\kk)$, $\mathfrak{g}_{zzzz}^\mathrm{orb}(\kk)$ can be calculated in an analogous manner. 
In particular, the functional form of $m_z^\mathrm{orb}$ coincides with the one of $m_z^\mathrm{spin}$ up to the leading order of $k_z$. 
In this case, $\mathfrak{g}_{zzzz}^\mathrm{orb}(\kk)$ is obtained by simply replacing $m_z^\mathrm{spin}$ with $m_z^\mathrm{orb}$ in the expression for $\mathfrak{g}_{zzzz}^\mathrm{spin}(\kk)$. 
However, $\kk$--dependent suppression factors $\varDelta E_\mathrm{vc/vh}$ of $m_z^\mathrm{orb}$ are not accurately taken into account if one retains only terms up to the leading order of $k_z$. A reasonable extension which integrates such higher-order terms of $\kk$ and converges to the expression above in the limit that $|\kk| \to 0$ is by substituting for $m_z^\mathrm{orb}$ and $v_z$ without the usage of Taylor series. This procedure was taken in the main text. 

If a system temperature $T$ is low enough that elastic scattering is still dominant, an extension from the absolute zero temperature to such a low temperature range may be performed by replacing $E_\mathrm{F}$ with $E_\mathrm{v}(\kk)$ in $\tau_\kk^\mathrm{F}$, and allowing $f_0$ to be $T$--dependent. By writing $\tau_\kk$ for the resulting effective relaxation time, we have
\begin{align}
    \gzzzz^\mathrm{orb/spin} &= 
    \int\frac{\mathrm{d}\kk}{(2\pi)^3}
    \mathfrak{g}_{zzzz}^\mathrm{orb/spin}(\kk)
    \left.\left(-\frac{\partial f_0}{\partial E}\right)\right|_{E=E_\mathrm{v}}, \\
    \mathfrak{g}_{zzzz}^\mathrm{orb/spin}(\kk) 
    &=-\frac{2}{5}(\tau_\kk)^2\frac{e^3}{\hbar^2}
    \frac{\partial}{\partial k_z}
    \left[
        (v_z)^2\frac{\partial}{\partial k_z}
        \frac{m_z^\mathrm{orb/spin}}{v_z}
    \right].
\end{align}

\subsection{Calculated carrier-density dependences of nonlinear electrical conductivities due to orbital and spin magnetic moments}
In Fig.~S6, we show the carrier-density dependences of $\gzzzz^\mathrm{orb}$ and $\gzzzz^\mathrm{spin}$ at 20 K. Both conductivities are normalized by respective maximum/minimum values. A peak structure is more clearly visible in $\gzzzz^\mathrm{orb}$ than $\gzzzz^\mathrm{spin}$. Because the only difference between normalized $\gzzzz^\mathrm{orb}$ and $\gzzzz^\mathrm{spin}$ is the functional form of the magnetic moment, the peak structure of $\gzzzz^\mathrm{orb}$ is attributed mainly to the energy dependence of the orbital magnetic moment, which is concentrated around the valence band top.

\renewcommand{\thefigure}{S6}
\begin{figure}[ht]
\centering
        \includegraphics[width=65mm]{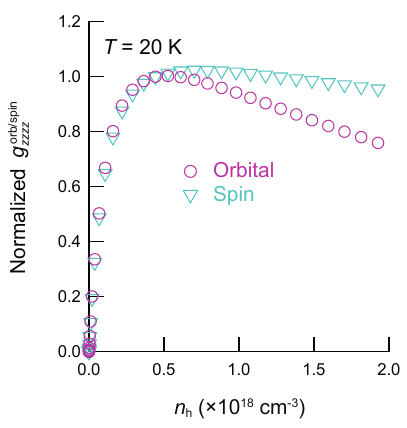}
\caption{
Calculated carrier-density $n_\mathrm{h}$ dependences of normalized nonlinear conductances due to orbital ($\gzzzz^\mathrm{orb}$) and spin ($\gzzzz^\mathrm{spin}$) magnetic moments of Bloch electrons. The temperature is set to 20 K.
}
\end{figure}


\begin{thebibliography}{53}%
\makeatletter
\providecommand \@ifxundefined [1]{%
 \@ifx{#1\undefined}
}%
\providecommand \@ifnum [1]{%
 \ifnum #1\expandafter \@firstoftwo
 \else \expandafter \@secondoftwo
 \fi
}%
\providecommand \@ifx [1]{%
 \ifx #1\expandafter \@firstoftwo
 \else \expandafter \@secondoftwo
 \fi
}%
\providecommand \natexlab [1]{#1}%
\providecommand \enquote  [1]{``#1''}%
\providecommand \bibnamefont  [1]{#1}%
\providecommand \bibfnamefont [1]{#1}%
\providecommand \citenamefont [1]{#1}%
\providecommand \href@noop [0]{\@secondoftwo}%
\providecommand \href [0]{\begingroup \@sanitize@url \@href}%
\providecommand \@href[1]{\@@startlink{#1}\@@href}%
\providecommand \@@href[1]{\endgroup#1\@@endlink}%
\providecommand \@sanitize@url [0]{\catcode `\\12\catcode `\$12\catcode `\&12\catcode `\#12\catcode `\^12\catcode `\_12\catcode `\%12\relax}%
\providecommand \@@startlink[1]{}%
\providecommand \@@endlink[0]{}%
\providecommand \url  [0]{\begingroup\@sanitize@url \@url }%
\providecommand \@url [1]{\endgroup\@href {#1}{\urlprefix }}%
\providecommand \urlprefix  [0]{URL }%
\providecommand \Eprint [0]{\href }%
\providecommand \doibase [0]{https://doi.org/}%
\providecommand \selectlanguage [0]{\@gobble}%
\providecommand \bibinfo  [0]{\@secondoftwo}%
\providecommand \bibfield  [0]{\@secondoftwo}%
\providecommand \translation [1]{[#1]}%
\providecommand \BibitemOpen [0]{}%
\providecommand \bibitemStop [0]{}%
\providecommand \bibitemNoStop [0]{.\EOS\space}%
\providecommand \EOS [0]{\spacefactor3000\relax}%
\providecommand \BibitemShut  [1]{\csname bibitem#1\endcsname}%
\let\auto@bib@innerbib\@empty
\bibitem [{\citenamefont {Rikken}\ \emph {et~al.}(2001)\citenamefont {Rikken}, \citenamefont {F{\"o}lling},\ and\ \citenamefont {Wyder}}]{RikkenPRL2001}%
  \BibitemOpen
  \bibfield  {author} {\bibinfo {author} {\bibfnamefont {G.~L. J.~A.}\ \bibnamefont {Rikken}}, \bibinfo {author} {\bibfnamefont {J.}~\bibnamefont {F{\"o}lling}},\ and\ \bibinfo {author} {\bibfnamefont {P.}~\bibnamefont {Wyder}},\ }\href@noop {} {\bibfield  {journal} {\bibinfo {title} {Electrical Magnetochiral Anisotropy.}} {\bibinfo  {journal} {Phys. Rev. Lett}\ }\textbf {\bibinfo {volume} {87}},\ \bibinfo {pages} {236602} (\bibinfo {year} {2001})}\BibitemShut {NoStop}%
\bibitem [{\citenamefont {Rikken}\ and\ \citenamefont {Wyder}(2005)}]{RikkenPRL2005}%
  \BibitemOpen
  \bibfield  {author} {\bibinfo {author} {\bibfnamefont {G.~L. J.~A.}\ \bibnamefont {Rikken}}\ and\ \bibinfo {author} {\bibfnamefont {P.}~\bibnamefont {Wyder}},\ }\href@noop {} {\bibfield  {journal} {\bibinfo {title} {Magnetoelectric Anisotropy in Diffusive Transport.}} {\bibinfo  {journal} {Phys. Rev. Lett}\ }\textbf {\bibinfo {volume} {94}},\ \bibinfo {pages} {016601} (\bibinfo {year} {2005})}\BibitemShut {NoStop}%
\bibitem [{\citenamefont {Tokura}\ and\ \citenamefont {Nagaosa}(2018)}]{TokuraNatCom2018}%
  \BibitemOpen
  \bibfield  {author} {\bibinfo {author} {\bibfnamefont {Y.}~\bibnamefont {Tokura}}\ and\ \bibinfo {author} {\bibfnamefont {N.}~\bibnamefont {Nagaosa}},\ }\href@noop {} {\bibfield  {journal} {\bibinfo {title} {Nonreciprocal responses from non-centrosymmetric quantum materials.}} {\bibinfo  {journal} {Nat. Commun.}\ }\textbf {\bibinfo {volume} {9}},\ \bibinfo {pages} {3740} (\bibinfo {year} {2018})}\BibitemShut {NoStop}%
\bibitem [{\citenamefont {Ideue}\ \emph {et~al.}(2017)\citenamefont {Ideue}, \citenamefont {Hamamoto}, \citenamefont {Koshikawa}, \citenamefont {Ezawa}, \citenamefont {Shimizu}, \citenamefont {Kaneko}, \citenamefont {Tokura}, \citenamefont {Nagaosa},\ and\ \citenamefont {Iwasa}}]{ideueNatPhys2017}%
  \BibitemOpen
  \bibfield  {author} {\bibinfo {author} {\bibfnamefont {T.}~\bibnamefont {Ideue}}, \bibinfo {author} {\bibfnamefont {K.}~\bibnamefont {Hamamoto}}, \bibinfo {author} {\bibfnamefont {S.}~\bibnamefont {Koshikawa}}, \bibinfo {author} {\bibfnamefont {M.}~\bibnamefont {Ezawa}}, \bibinfo {author} {\bibfnamefont {S.}~\bibnamefont {Shimizu}}, \bibinfo {author} {\bibfnamefont {Y.}~\bibnamefont {Kaneko}}, \bibinfo {author} {\bibfnamefont {Y.}~\bibnamefont {Tokura}}, \bibinfo {author} {\bibfnamefont {N.}~\bibnamefont {Nagaosa}},\ and\ \bibinfo {author} {\bibfnamefont {Y.}~\bibnamefont {Iwasa}},\ }\href@noop {} {\bibfield  {journal} {\bibinfo {title} {Bulk rectification effect in a polar semiconductor.}} {\bibinfo  {journal} {Nat. Phys.}\ }\textbf {\bibinfo {volume} {13}},\ \bibinfo {pages} {578} (\bibinfo {year} {2017})}\BibitemShut {NoStop}%
\bibitem [{\citenamefont {Pop}\ \emph {et~al.}(2014)\citenamefont {Pop}, \citenamefont {Auban-Senzier}, \citenamefont {Canadell}, \citenamefont {Rikken},\ and\ \citenamefont {Avarvari}}]{PopNatCommun2014}%
  \BibitemOpen
  \bibfield  {author} {\bibinfo {author} {\bibfnamefont {F.}~\bibnamefont {Pop}}, \bibinfo {author} {\bibfnamefont {P.}~\bibnamefont {Auban-Senzier}}, \bibinfo {author} {\bibfnamefont {E.}~\bibnamefont {Canadell}}, \bibinfo {author} {\bibfnamefont {G.~L. J.~A.}\ \bibnamefont {Rikken}},\ and\ \bibinfo {author} {\bibfnamefont {N.}~\bibnamefont {Avarvari}},\ }\href@noop {} {\bibfield  {journal} {\bibinfo {title} {Electrical magnetochiral anisotropy in a bulk chiral molecular conductor.}} {\bibinfo  {journal} {Nat. Commun.}\ }\textbf {\bibinfo {volume} {5}},\ \bibinfo {pages} {3757} (\bibinfo {year} {2014})}\BibitemShut {NoStop}%
\bibitem [{\citenamefont {Yokouchi}\ \emph {et~al.}(2017)\citenamefont {Yokouchi}, \citenamefont {Kanazawa}, \citenamefont {Kikkawa}, \citenamefont {Morikawa}, \citenamefont {Shibata}, \citenamefont {Arima}, \citenamefont {Taguchi}, \citenamefont {Kagawa},\ and\ \citenamefont {Tokura}}]{YokouchiPRL2017}%
  \BibitemOpen
  \bibfield  {author} {\bibinfo {author} {\bibfnamefont {T.}~\bibnamefont {Yokouchi}}, \bibinfo {author} {\bibfnamefont {N.}~\bibnamefont {Kanazawa}}, \bibinfo {author} {\bibfnamefont {A.}~\bibnamefont {Kikkawa}}, \bibinfo {author} {\bibfnamefont {D.}~\bibnamefont {Morikawa}}, \bibinfo {author} {\bibfnamefont {K.}~\bibnamefont {Shibata}}, \bibinfo {author} {\bibfnamefont {T.}~\bibnamefont {Arima}}, \bibinfo {author} {\bibfnamefont {Y.}~\bibnamefont {Taguchi}}, \bibinfo {author} {\bibfnamefont {F.}~\bibnamefont {Kagawa}},\ and\ \bibinfo {author} {\bibfnamefont {Y.}~\bibnamefont {Tokura}},\ }\href@noop {} {\bibfield  {journal} {\bibinfo {title} {Electrical magnetochiral effect induced by chiral spin fluctuations.}} {\bibinfo  {journal} {Nat. Commun.}\ }\textbf {\bibinfo {volume} {8}},\ \bibinfo {pages} {866} (\bibinfo {year} {2017})}\BibitemShut {NoStop}%
\bibitem [{\citenamefont {Avci}\ \emph {et~al.}(2015)\citenamefont {Avci}, \citenamefont {Garello}, \citenamefont {Ghosh}, \citenamefont {Gabureac}, \citenamefont {Alvarado},\ and\ \citenamefont {Gambardella}}]{AvciNatPhys2015}%
  \BibitemOpen
  \bibfield  {author} {\bibinfo {author} {\bibfnamefont {C.}~\bibnamefont {Avci}}, \bibinfo {author} {\bibfnamefont {K.}~\bibnamefont {Garello}}, \bibinfo {author} {\bibfnamefont {A.}~\bibnamefont {Ghosh}}, \bibinfo {author} {\bibfnamefont {M.}~\bibnamefont {Gabureac}}, \bibinfo {author} {\bibfnamefont {S.~F.}\ \bibnamefont {Alvarado}},\ and\ \bibinfo {author} {\bibfnamefont {P.}~\bibnamefont {Gambardella}},\ }\href@noop {} {\bibfield  {journal} {\bibinfo {title} {Unidirectional spin Hall magnetoresistance in ferromagnet/normal metal bilayers.}} {\bibinfo  {journal} {Nat. Phys.}\ }\textbf {\bibinfo {volume} {11}},\ \bibinfo {pages} {570} (\bibinfo {year} {2015})}\BibitemShut {NoStop}%
\bibitem [{\citenamefont {He}\ \emph {et~al.}(2019)\citenamefont {He}, \citenamefont {Hsu}, \citenamefont {Shi}, \citenamefont {Cai}, \citenamefont {Wang}, \citenamefont {Wang}, \citenamefont {Eda}, \citenamefont {Lin}, \citenamefont {Pereira},\ and\ \citenamefont {Yang}}]{HeNatCommun2019}%
  \BibitemOpen
  \bibfield  {author} {\bibinfo {author} {\bibfnamefont {P.}~\bibnamefont {He}}, \bibinfo {author} {\bibfnamefont {C.-H.}\ \bibnamefont {Hsu}}, \bibinfo {author} {\bibfnamefont {S.}~\bibnamefont {Shi}}, \bibinfo {author} {\bibfnamefont {K.}~\bibnamefont {Cai}}, \bibinfo {author} {\bibfnamefont {J.}~\bibnamefont {Wang}}, \bibinfo {author} {\bibfnamefont {Q.}~\bibnamefont {Wang}}, \bibinfo {author} {\bibfnamefont {G.}~\bibnamefont {Eda}}, \bibinfo {author} {\bibfnamefont {H.}~\bibnamefont {Lin}}, \bibinfo {author} {\bibfnamefont {V.}~\bibnamefont {Pereira}},\ and\ \bibinfo {author} {\bibfnamefont {H.}~\bibnamefont {Yang}},\ }\href@noop {} {\bibfield  {journal} {\bibinfo {title} {Nonlinear magnetotransport shaped by Fermi surface topology and convexity.}} {\bibinfo  {journal} {Nat. Commun.}\ }\textbf {\bibinfo {volume} {10}},\ \bibinfo {pages} {1290} (\bibinfo {year} {2019})}\BibitemShut {NoStop}%
\bibitem [{\citenamefont {Yokouchi}\ \emph {et~al.}(2023)\citenamefont {Yokouchi}, \citenamefont {Ikeda}, \citenamefont {Morimoto},\ and\ \citenamefont {Shiomi}}]{YokouchiPRL2023}%
  \BibitemOpen
  \bibfield  {author} {\bibinfo {author} {\bibfnamefont {T.}~\bibnamefont {Yokouchi}}, \bibinfo {author} {\bibfnamefont {Y.}~\bibnamefont {Ikeda}}, \bibinfo {author} {\bibfnamefont {T.}~\bibnamefont {Morimoto}},\ and\ \bibinfo {author} {\bibfnamefont {Y.}~\bibnamefont {Shiomi}},\ }\href@noop {} {\bibfield  {journal} {\bibinfo {title} {Giant Magnetochiral Anisotropi in Weyl Semimetal WTe$_2$ Induced by Diverging Berry Curvature.}} {\bibinfo  {journal} {Phys. Rev. Lett.}\ }\textbf {\bibinfo {volume} {130}},\ \bibinfo {pages} {136301} (\bibinfo {year} {2023})}\BibitemShut {NoStop}%
\bibitem [{\citenamefont {Chang}\ and\ \citenamefont {Niu}(1996)}]{ChangPhysRevB1996}%
  \BibitemOpen
  \bibfield  {author} {\bibinfo {author} {\bibfnamefont {M.-C.}\ \bibnamefont {Chang}}\ and\ \bibinfo {author} {\bibfnamefont {Q.}~\bibnamefont {Niu}},\ }\href@noop {} {\bibfield  {journal} {\bibinfo {title} {Berry phase, hyperorbits, and the Hofstadter spectrum: Semiclassical dynamics in magnetic Bloch bands.}} {\bibinfo  {journal} {Phys. Rev. B}\ }\textbf {\bibinfo {volume} {53}},\ \bibinfo {pages} {7010} (\bibinfo {year} {1996})}\BibitemShut {NoStop}%
\bibitem [{\citenamefont {Xiao}\ \emph {et~al.}(2010)\citenamefont {Xiao}, \citenamefont {Chang},\ and\ \citenamefont {Niu}}]{XiaoRevModPhys2010}%
  \BibitemOpen
  \bibfield  {author} {\bibinfo {author} {\bibfnamefont {D.}~\bibnamefont {Xiao}}, \bibinfo {author} {\bibfnamefont {M.-C.}\ \bibnamefont {Chang}},\ and\ \bibinfo {author} {\bibfnamefont {Q.}~\bibnamefont {Niu}},\ }\href@noop {} {\bibfield  {journal} {\bibinfo {title} {Berry phase effects on electronic properties.}} {\bibinfo  {journal} {Rev. Mod. Phys.}\ }\textbf {\bibinfo {volume} {82}},\ \bibinfo {pages} {1959} (\bibinfo {year} {2010})}\BibitemShut {NoStop}%
\bibitem [{\citenamefont {Yoda}\ \emph {et~al.}(2015)\citenamefont {Yoda}, \citenamefont {Yokoyama},\ and\ \citenamefont {Murakami}}]{YodaSciRep2015}%
  \BibitemOpen
  \bibfield  {author} {\bibinfo {author} {\bibfnamefont {T.}~\bibnamefont {Yoda}}, \bibinfo {author} {\bibfnamefont {Y.}~\bibnamefont {Yokoyama}},\ and\ \bibinfo {author} {\bibfnamefont {S.}~\bibnamefont {Murakami}},\ }\href@noop {} {\bibfield  {journal} {\bibinfo {title} {Current-induced Orbital and Spin Magnetizations in Crystals with Helical Structure.}} {\bibinfo  {journal} {Sci. Rep.}\ }\textbf {\bibinfo {volume} {5}},\ \bibinfo {pages} {12024} (\bibinfo {year} {2015})}\BibitemShut {NoStop}%
\bibitem [{\citenamefont {Zhong}\ \emph {et~al.}(2016)\citenamefont {Zhong}, \citenamefont {Moore},\ and\ \citenamefont {Souza}}]{ZhongPRL2016}%
  \BibitemOpen
  \bibfield  {author} {\bibinfo {author} {\bibfnamefont {S.}~\bibnamefont {Zhong}}, \bibinfo {author} {\bibfnamefont {J.~E.}\ \bibnamefont {Moore}},\ and\ \bibinfo {author} {\bibfnamefont {I.}~\bibnamefont {Souza}},\ }\href@noop {} {\bibfield  {journal} {\bibinfo {title} {Current-induced Orbital and Spin Magnetizations in Crystals with Helical Structure.}} {\bibinfo  {journal} {Phys. Rev. Lett.}\ }\textbf {\bibinfo {volume} {116}},\ \bibinfo {pages} {077201} (\bibinfo {year} {2016})}\BibitemShut {NoStop}%
\bibitem [{\citenamefont {Yoda}\ \emph {et~al.}(2018)\citenamefont {Yoda}, \citenamefont {Yokoyama},\ and\ \citenamefont {Murakami}}]{YodaNanoLett2018}%
  \BibitemOpen
  \bibfield  {author} {\bibinfo {author} {\bibfnamefont {T.}~\bibnamefont {Yoda}}, \bibinfo {author} {\bibfnamefont {Y.}~\bibnamefont {Yokoyama}},\ and\ \bibinfo {author} {\bibfnamefont {S.}~\bibnamefont {Murakami}},\ }\href@noop {} {\bibfield  {journal} {\bibinfo {title} {Orbital Edelstein Effect as a Condensed-Matter Analog of Solenoids.}} {\bibinfo  {journal} {Nano Lett.}\ }\textbf {\bibinfo {volume} {18}},\ \bibinfo {pages} {916} (\bibinfo {year} {2018})}\BibitemShut {NoStop}%
\bibitem [{\citenamefont {He}\ \emph {et~al.}(2020)\citenamefont {He}, \citenamefont {Goldhaber-Gordon},\ and\ \citenamefont {Law}}]{HeNatCommun2020}%
  \BibitemOpen
  \bibfield  {author} {\bibinfo {author} {\bibfnamefont {W.-Y.}\ \bibnamefont {He}}, \bibinfo {author} {\bibfnamefont {D.}~\bibnamefont {Goldhaber-Gordon}},\ and\ \bibinfo {author} {\bibfnamefont {K.~T.}\ \bibnamefont {Law}},\ }\href@noop {} {\bibfield  {journal} {\bibinfo {title} {Giant orbital magnetoelectric effect and current-induced magnetization switching in twisted bilayer graphene.}} {\bibinfo  {journal} {Nat. Commun.}\ }\textbf {\bibinfo {volume} {11}},\ \bibinfo {pages} {1650} (\bibinfo {year} {2020})}\BibitemShut {NoStop}%
\bibitem [{\citenamefont {He}\ and\ \citenamefont {Law}(2021)}]{HePhysRevRes2021}%
  \BibitemOpen
  \bibfield  {author} {\bibinfo {author} {\bibfnamefont {W.-Y.}\ \bibnamefont {He}}\ and\ \bibinfo {author} {\bibfnamefont {K.~T.}\ \bibnamefont {Law}},\ }\href@noop {} {\bibfield  {journal} {\bibinfo {title} {Superconducting orbital magnetoelectric effect and its evolution across the superconductor-normal metal phase transition.}} {\bibinfo  {journal} {Phys. Rev. Res.}\ }\textbf {\bibinfo {volume} {3}},\ \bibinfo {pages} {L032012} (\bibinfo {year} {2021})}\BibitemShut {NoStop}%
\bibitem [{\citenamefont {Chirolli}\ \emph {et~al.}(2022)\citenamefont {Chirolli}, \citenamefont {Mercaldo}, \citenamefont {Guarcello}, \citenamefont {Giazotto},\ and\ \citenamefont {Cuoco}}]{ChirolliPRL2022}%
  \BibitemOpen
  \bibfield  {author} {\bibinfo {author} {\bibfnamefont {L.}~\bibnamefont {Chirolli}}, \bibinfo {author} {\bibfnamefont {M.~T.}\ \bibnamefont {Mercaldo}}, \bibinfo {author} {\bibfnamefont {C.}~\bibnamefont {Guarcello}}, \bibinfo {author} {\bibfnamefont {F.}~\bibnamefont {Giazotto}},\ and\ \bibinfo {author} {\bibfnamefont {M.}~\bibnamefont {Cuoco}},\ }\href@noop {} {\bibfield  {journal} {\bibinfo {title} {Colossal Orbital Edelstein Effect in Noncentrosymmetric Superconductors.}} {\bibinfo  {journal} {Phys. Rev. Lett.}\ }\textbf {\bibinfo {volume} {128}},\ \bibinfo {pages} {217703} (\bibinfo {year} {2022})}\BibitemShut {NoStop}%
\bibitem [{\citenamefont {Liu}\ \emph {et~al.}(2023)\citenamefont {Liu}, \citenamefont {Souza},\ and\ \citenamefont {Tsirkin}}]{liu2023electrical}%
  \BibitemOpen
  \bibfield  {author} {\bibinfo {author} {\bibfnamefont {X.}~\bibnamefont {Liu}}, \bibinfo {author} {\bibfnamefont {I.}~\bibnamefont {Souza}},\ and\ \bibinfo {author} {\bibfnamefont {S.~S.}\ \bibnamefont {Tsirkin}},\ }\href@noop {} {\bibfield  {journal} {\bibinfo {title} {Electrical magnetochiral anisotropy in trigonal tellurium from first principles.}} {\bibinfo  {journal} {arXiv preprint arXiv:2303.10164}\ } (\bibinfo {year} {2023})}\BibitemShut {NoStop}%
\bibitem [{\citenamefont {Zhu}\ \emph {et~al.}(2017)\citenamefont {Zhu}, \citenamefont {Cai}, \citenamefont {Yi}, \citenamefont {Chen}, \citenamefont {Dai}, \citenamefont {Niu}, \citenamefont {Guo}, \citenamefont {Xie}, \citenamefont {Liu}, \citenamefont {Cho}, \citenamefont {Jia},\ and\ \citenamefont {Zhang}}]{ZhuPRL2017}%
  \BibitemOpen
  \bibfield  {author} {\bibinfo {author} {\bibfnamefont {Z.}~\bibnamefont {Zhu}}, \bibinfo {author} {\bibfnamefont {X.}~\bibnamefont {Cai}}, \bibinfo {author} {\bibfnamefont {S.}~\bibnamefont {Yi}}, \bibinfo {author} {\bibfnamefont {J.}~\bibnamefont {Chen}}, \bibinfo {author} {\bibfnamefont {Y.}~\bibnamefont {Dai}}, \bibinfo {author} {\bibfnamefont {C.}~\bibnamefont {Niu}}, \bibinfo {author} {\bibfnamefont {Z.}~\bibnamefont {Guo}}, \bibinfo {author} {\bibfnamefont {M.}~\bibnamefont {Xie}}, \bibinfo {author} {\bibfnamefont {F.}~\bibnamefont {Liu}}, \bibinfo {author} {\bibfnamefont {J.-H.}\ \bibnamefont {Cho}}, \bibinfo {author} {\bibfnamefont {Y.}~\bibnamefont {Jia}},\ and\ \bibinfo {author} {\bibfnamefont {Z.}~\bibnamefont {Zhang}},\ }\href@noop {} {\bibfield  {journal} {\bibinfo {title} {Multivalency-Driven Formation of Te-Based Monolayer Materials: A Combined First-Principles and Experimental study.}} {\bibinfo  {journal} {Phys. Rev. Lett.}\ }\textbf {\bibinfo {volume} {119}},\ \bibinfo {pages} {106101} (\bibinfo {year} {2017})}\BibitemShut {NoStop}%
\bibitem [{\citenamefont {Yi}\ \emph {et~al.}(2018)\citenamefont {Yi}, \citenamefont {Zhu}, \citenamefont {Cai}, \citenamefont {Jia},\ and\ \citenamefont {Cho}}]{YiInorgChem2018}%
  \BibitemOpen
  \bibfield  {author} {\bibinfo {author} {\bibfnamefont {S.}~\bibnamefont {Yi}}, \bibinfo {author} {\bibfnamefont {Z.}~\bibnamefont {Zhu}}, \bibinfo {author} {\bibfnamefont {X.}~\bibnamefont {Cai}}, \bibinfo {author} {\bibfnamefont {Y.}~\bibnamefont {Jia}},\ and\ \bibinfo {author} {\bibfnamefont {J.-H.}\ \bibnamefont {Cho}},\ }\href@noop {} {\bibfield  {journal} {\bibinfo {title} {The Nature of Bonding in Bulk Tellurium Composed of OneDimensional Helical Chains.}} {\bibinfo  {journal} {Inorg. Chem.}\ }\textbf {\bibinfo {volume} {57}},\ \bibinfo {pages} {5083} (\bibinfo {year} {2018})}\BibitemShut {NoStop}%
\bibitem [{\citenamefont {Sakano}\ \emph {et~al.}(2020)\citenamefont {Sakano}, \citenamefont {Hirayama}, \citenamefont {Takahashi}, \citenamefont {Akebi}, \citenamefont {Nakayama}, \citenamefont {Kuroda}, \citenamefont {Taguchi}, \citenamefont {Yoshikawa}, \citenamefont {Miyamoto}, \citenamefont {Okuda}, \citenamefont {Ono}, \citenamefont {Kumigashira}, \citenamefont {Ideue}, \citenamefont {Iwasa}, \citenamefont {Mitsuishi}, \citenamefont {Ishizaka}, \citenamefont {Shin}, \citenamefont {Miyake}, \citenamefont {Murakami}, \citenamefont {Sasagawa},\ and\ \citenamefont {Kondo}}]{SakanoPRL2020}%
  \BibitemOpen
  \bibfield  {author} {\bibinfo {author} {\bibfnamefont {M.}~\bibnamefont {Sakano}}, \bibinfo {author} {\bibfnamefont {M.}~\bibnamefont {Hirayama}}, \bibinfo {author} {\bibfnamefont {T.}~\bibnamefont {Takahashi}}, \bibinfo {author} {\bibfnamefont {S.}~\bibnamefont {Akebi}}, \bibinfo {author} {\bibfnamefont {M.}~\bibnamefont {Nakayama}}, \bibinfo {author} {\bibfnamefont {K.}~\bibnamefont {Kuroda}}, \bibinfo {author} {\bibfnamefont {K.}~\bibnamefont {Taguchi}}, \bibinfo {author} {\bibfnamefont {T.}~\bibnamefont {Yoshikawa}}, \bibinfo {author} {\bibfnamefont {K.}~\bibnamefont {Miyamoto}}, \bibinfo {author} {\bibfnamefont {T.}~\bibnamefont {Okuda}}, \bibinfo {author} {\bibfnamefont {K.}~\bibnamefont {Ono}}, \bibinfo {author} {\bibfnamefont {H.}~\bibnamefont {Kumigashira}}, \bibinfo {author} {\bibfnamefont {T.}~\bibnamefont {Ideue}}, \bibinfo {author} {\bibfnamefont {Y.}~\bibnamefont {Iwasa}}, \bibinfo {author} {\bibfnamefont {N.}~\bibnamefont {Mitsuishi}}, \bibinfo {author} {\bibfnamefont {K.}~\bibnamefont
  {Ishizaka}}, \bibinfo {author} {\bibfnamefont {S.}~\bibnamefont {Shin}}, \bibinfo {author} {\bibfnamefont {T.}~\bibnamefont {Miyake}}, \bibinfo {author} {\bibfnamefont {S.}~\bibnamefont {Murakami}}, \bibinfo {author} {\bibfnamefont {T.}~\bibnamefont {Sasagawa}},\ and\ \bibinfo {author} {\bibfnamefont {T.}~\bibnamefont {Kondo}},\ }\href@noop {} {\bibfield  {journal} {\bibinfo {title} {Radial Spin Texture in Elemental Tellurium with Chiral Crystal Structure.}} {\bibinfo  {journal} {Phys. Rev. Lett.}\ }\textbf {\bibinfo {volume} {124}},\ \bibinfo {pages} {136404} (\bibinfo {year} {2020})}\BibitemShut {NoStop}%
\bibitem [{\citenamefont {Gatti}\ \emph {et~al.}(2020)\citenamefont {Gatti}, \citenamefont {Gosálbez-Martínez}, \citenamefont {Tsirkin}, \citenamefont {Fanciulli}, \citenamefont {Puppin}, \citenamefont {Polishchuk}, \citenamefont {Moser}, \citenamefont {Testa}, \citenamefont {Martino}, \citenamefont {Roth}, \citenamefont {Bugnon}, \citenamefont {Moreschini}, \citenamefont {Bostwick}, \citenamefont {Jozwiak}, \citenamefont {Rotenberg}, \citenamefont {Santo}, \citenamefont {Petaccia}, \citenamefont {Vobornik}, \citenamefont {Fujii}, \citenamefont {Wong}, \citenamefont {Jariwala}, \citenamefont {Atwater}, \citenamefont {Rønnow}, \citenamefont {Chergui}, \citenamefont {Yazyev}, \citenamefont {Grioni},\ and\ \citenamefont {Crepaldi}}]{GattiPRL2020}%
  \BibitemOpen
  \bibfield  {author} {\bibinfo {author} {\bibfnamefont {G.}~\bibnamefont {Gatti}}, \bibinfo {author} {\bibfnamefont {D.}~\bibnamefont {Gosálbez-Martínez}}, \bibinfo {author} {\bibfnamefont {S.~S.}\ \bibnamefont {Tsirkin}}, \bibinfo {author} {\bibfnamefont {M.}~\bibnamefont {Fanciulli}}, \bibinfo {author} {\bibfnamefont {M.}~\bibnamefont {Puppin}}, \bibinfo {author} {\bibfnamefont {S.}~\bibnamefont {Polishchuk}}, \bibinfo {author} {\bibfnamefont {S.}~\bibnamefont {Moser}}, \bibinfo {author} {\bibfnamefont {L.}~\bibnamefont {Testa}}, \bibinfo {author} {\bibfnamefont {E.}~\bibnamefont {Martino}}, \bibinfo {author} {\bibfnamefont {S.}~\bibnamefont {Roth}}, \bibinfo {author} {\bibfnamefont {P.}~\bibnamefont {Bugnon}}, \bibinfo {author} {\bibfnamefont {L.}~\bibnamefont {Moreschini}}, \bibinfo {author} {\bibfnamefont {A.}~\bibnamefont {Bostwick}}, \bibinfo {author} {\bibfnamefont {C.}~\bibnamefont {Jozwiak}}, \bibinfo {author} {\bibfnamefont {E.}~\bibnamefont {Rotenberg}}, \bibinfo {author} {\bibfnamefont
  {G.~D.}\ \bibnamefont {Santo}}, \bibinfo {author} {\bibfnamefont {L.}~\bibnamefont {Petaccia}}, \bibinfo {author} {\bibfnamefont {I.}~\bibnamefont {Vobornik}}, \bibinfo {author} {\bibfnamefont {J.}~\bibnamefont {Fujii}}, \bibinfo {author} {\bibfnamefont {J.}~\bibnamefont {Wong}}, \bibinfo {author} {\bibfnamefont {D.}~\bibnamefont {Jariwala}}, \bibinfo {author} {\bibfnamefont {H.~A.}\ \bibnamefont {Atwater}}, \bibinfo {author} {\bibfnamefont {H.~M.}\ \bibnamefont {Rønnow}}, \bibinfo {author} {\bibfnamefont {M.}~\bibnamefont {Chergui}}, \bibinfo {author} {\bibfnamefont {O.}~\bibnamefont {Yazyev}}, \bibinfo {author} {\bibfnamefont {M.}~\bibnamefont {Grioni}},\ and\ \bibinfo {author} {\bibfnamefont {A.}~\bibnamefont {Crepaldi}},\ }\href@noop {} {\bibfield  {journal} {\bibinfo {title} {Radial Spin Texture of the Weyl Fermions in Chiral Tellurium.}} {\bibinfo  {journal} {Phys. Rev. Lett.}\ }\textbf {\bibinfo {volume} {125}},\ \bibinfo {pages} {216402} (\bibinfo {year} {2020})}\BibitemShut {NoStop}%
\bibitem [{\citenamefont {Nomura}(1960)}]{NomuraPRL1960}%
  \BibitemOpen
  \bibfield  {author} {\bibinfo {author} {\bibfnamefont {K.~C.}\ \bibnamefont {Nomura}},\ }\href@noop {} {\bibfield  {journal} {\bibinfo {title} {Optical Activity in Tellurium.}} {\bibinfo  {journal} {Phys.Rev. Lett.}\ }\textbf {\bibinfo {volume} {5}},\ \bibinfo {pages} {500} (\bibinfo {year} {1960})}\BibitemShut {NoStop}%
\bibitem [{\citenamefont {Fukuda}\ \emph {et~al.}(1975)\citenamefont {Fukuda}, \citenamefont {Shiosaki},\ and\ \citenamefont {Kawabata}}]{FukudaPSS1975}%
  \BibitemOpen
  \bibfield  {author} {\bibinfo {author} {\bibfnamefont {S.}~\bibnamefont {Fukuda}}, \bibinfo {author} {\bibfnamefont {T.}~\bibnamefont {Shiosaki}},\ and\ \bibinfo {author} {\bibfnamefont {A.}~\bibnamefont {Kawabata}},\ }\href@noop {} {\bibfield  {journal} {\bibinfo {title} {Infrared optical activity in tellurium.}} {\bibinfo  {journal} {Phys. Status Solidi (b)}\ }\textbf {\bibinfo {volume} {68}},\ \bibinfo {pages} {K107} (\bibinfo {year} {1975})}\BibitemShut {NoStop}%
\bibitem [{\citenamefont {Ades}\ and\ \citenamefont {Champness}(1975)}]{AdesJOSA1975}%
  \BibitemOpen
  \bibfield  {author} {\bibinfo {author} {\bibfnamefont {S.}~\bibnamefont {Ades}}\ and\ \bibinfo {author} {\bibfnamefont {C.~H.}\ \bibnamefont {Champness}},\ }\href@noop {} {\bibfield  {journal} {\bibinfo {title} {Optical activity of tellurium to 20 \textmu m.}} {\bibinfo  {journal} {J. Opt. Soc. Am.}\ }\textbf {\bibinfo {volume} {65}},\ \bibinfo {pages} {217} (\bibinfo {year} {1975})}\BibitemShut {NoStop}%
\bibitem [{\citenamefont {Stolze}\ \emph {et~al.}(1977)\citenamefont {Stolze}, \citenamefont {Lutz},\ and\ \citenamefont {Grosse}}]{StolzePSS1977}%
  \BibitemOpen
  \bibfield  {author} {\bibinfo {author} {\bibfnamefont {H.}~\bibnamefont {Stolze}}, \bibinfo {author} {\bibfnamefont {M.}~\bibnamefont {Lutz}},\ and\ \bibinfo {author} {\bibfnamefont {P.}~\bibnamefont {Grosse}},\ }\href@noop {} {\bibfield  {journal} {\bibinfo {title} {The optical activity of tellurium.}} {\bibinfo  {journal} {Phys. Status Solidi (b)}\ }\textbf {\bibinfo {volume} {82}},\ \bibinfo {pages} {457} (\bibinfo {year} {1977})}\BibitemShut {NoStop}%
\bibitem [{\citenamefont {Vorob'ev}\ \emph {et~al.}(1979)\citenamefont {Vorob'ev}, \citenamefont {Ivchenko}, \citenamefont {Pikus}, \citenamefont {Farbshtein}, \citenamefont {Shalygin},\ and\ \citenamefont {Shturbin}}]{VorobevPETF1979}%
  \BibitemOpen
  \bibfield  {author} {\bibinfo {author} {\bibfnamefont {L.~E.}\ \bibnamefont {Vorob'ev}}, \bibinfo {author} {\bibfnamefont {E.~L.}\ \bibnamefont {Ivchenko}}, \bibinfo {author} {\bibfnamefont {G.~E.}\ \bibnamefont {Pikus}}, \bibinfo {author} {\bibfnamefont {I.~I.}\ \bibnamefont {Farbshtein}}, \bibinfo {author} {\bibfnamefont {V.~A.}\ \bibnamefont {Shalygin}},\ and\ \bibinfo {author} {\bibfnamefont {A.~V.}\ \bibnamefont {Shturbin}},\ }\href@noop {} {\bibfield  {journal} {\bibinfo {title} {Optical activity in tellurium induced by a current.}} {\bibinfo  {journal} {Pis’ma Zh. Eksp. Teor. Fiz.}\ }\textbf {\bibinfo {volume} {29}},\ \bibinfo {pages} {485} (\bibinfo {year} {1979})}\BibitemShut {NoStop}%
\bibitem [{\citenamefont {Shalygin}\ \emph {et~al.}(2012)\citenamefont {Shalygin}, \citenamefont {Sofronov}, \citenamefont {Vorob’ev},\ and\ \citenamefont {Farbshtein}}]{ShalyginPSS2012}%
  \BibitemOpen
  \bibfield  {author} {\bibinfo {author} {\bibfnamefont {V.~A.}\ \bibnamefont {Shalygin}}, \bibinfo {author} {\bibfnamefont {A.~N.}\ \bibnamefont {Sofronov}}, \bibinfo {author} {\bibfnamefont {L.~E.}\ \bibnamefont {Vorob’ev}},\ and\ \bibinfo {author} {\bibfnamefont {I.~I.}\ \bibnamefont {Farbshtein}},\ }\href@noop {} {\bibfield  {journal} {\bibinfo {title} {Current-induced spin polarization of holes in tellurium.}} {\bibinfo  {journal} {Phys. Solid State}\ }\textbf {\bibinfo {volume} {54}},\ \bibinfo {pages} {2362} (\bibinfo {year} {2012})}\BibitemShut {NoStop}%
\bibitem [{\citenamefont {Furukawa}\ \emph {et~al.}(2017)\citenamefont {Furukawa}, \citenamefont {Shimokawa}, \citenamefont {Kobayashi},\ and\ \citenamefont {Itou}}]{FurukawaNatCom2017}%
  \BibitemOpen
  \bibfield  {author} {\bibinfo {author} {\bibfnamefont {T.}~\bibnamefont {Furukawa}}, \bibinfo {author} {\bibfnamefont {Y.}~\bibnamefont {Shimokawa}}, \bibinfo {author} {\bibfnamefont {K.}~\bibnamefont {Kobayashi}},\ and\ \bibinfo {author} {\bibfnamefont {T.}~\bibnamefont {Itou}},\ }\href@noop {} {\bibfield  {journal} {\bibinfo {title} {Observation of current-induced bulk magnetization in elemental tellurium.}} {\bibinfo  {journal} {Nat. Commun.}\ }\textbf {\bibinfo {volume} {8}},\ \bibinfo {pages} {954} (\bibinfo {year} {2017})}\BibitemShut {NoStop}%
\bibitem [{\citenamefont {Furukawa}\ \emph {et~al.}(2021)\citenamefont {Furukawa}, \citenamefont {Watanabe}, \citenamefont {Ogasawara}, \citenamefont {Kobayashi},\ and\ \citenamefont {Itou}}]{FurukawaPhysRevRes2021}%
  \BibitemOpen
  \bibfield  {author} {\bibinfo {author} {\bibfnamefont {T.}~\bibnamefont {Furukawa}}, \bibinfo {author} {\bibfnamefont {Y.}~\bibnamefont {Watanabe}}, \bibinfo {author} {\bibfnamefont {N.}~\bibnamefont {Ogasawara}}, \bibinfo {author} {\bibfnamefont {K.}~\bibnamefont {Kobayashi}},\ and\ \bibinfo {author} {\bibfnamefont {T.}~\bibnamefont {Itou}},\ }\href@noop {} {\bibfield  {journal} {\bibinfo {title} {Current-induced magnetization caused by crystal chirality in nonmagnetic elemental tellurium.}} {\bibinfo  {journal} {Phys. Rev. Res.}\ }\textbf {\bibinfo {volume} {3}},\ \bibinfo {pages} {023111} (\bibinfo {year} {2021})}\BibitemShut {NoStop}%
\bibitem [{\citenamefont {Koma}\ and\ \citenamefont {Tanaka}(1970)}]{koma1970etch}%
  \BibitemOpen
  \bibfield  {author} {\bibinfo {author} {\bibfnamefont {A.}~\bibnamefont {Koma}}\ and\ \bibinfo {author} {\bibfnamefont {S.}~\bibnamefont {Tanaka}},\ }\href@noop {} {\bibfield  {journal} {\bibinfo {title} {Etch pits and crystal structure of tellurium.}} {\bibinfo  {journal} {Phys. Stat. Sol.}\ }\textbf {\bibinfo {volume} {40}},\ \bibinfo {pages} {239} (\bibinfo {year} {1970})}\BibitemShut {NoStop}%
\bibitem [{\citenamefont {Calavalle}\ \emph {et~al.}(2022)\citenamefont {Calavalle}, \citenamefont {Suárez-Rodríguez}, \citenamefont {Martín-García}, \citenamefont {Johansson}, \citenamefont {Vaz}, \citenamefont {Yang}, \citenamefont {Maznichenko}, \citenamefont {Ostanin}, \citenamefont {Mateo-Alonso}, \citenamefont {Chuvilin}, \citenamefont {Mertig}, \citenamefont {Gobbi}, \citenamefont {Casanova},\ and\ \citenamefont {Hueso}}]{CalavalleNatMater2022}%
  \BibitemOpen
  \bibfield  {author} {\bibinfo {author} {\bibfnamefont {F.}~\bibnamefont {Calavalle}}, \bibinfo {author} {\bibfnamefont {M.}~\bibnamefont {Suárez-Rodríguez}}, \bibinfo {author} {\bibfnamefont {B.}~\bibnamefont {Martín-García}}, \bibinfo {author} {\bibfnamefont {A.}~\bibnamefont {Johansson}}, \bibinfo {author} {\bibfnamefont {D.~C.}\ \bibnamefont {Vaz}}, \bibinfo {author} {\bibfnamefont {H.}~\bibnamefont {Yang}}, \bibinfo {author} {\bibfnamefont {I.~V.}\ \bibnamefont {Maznichenko}}, \bibinfo {author} {\bibfnamefont {S.}~\bibnamefont {Ostanin}}, \bibinfo {author} {\bibfnamefont {A.}~\bibnamefont {Mateo-Alonso}}, \bibinfo {author} {\bibfnamefont {A.}~\bibnamefont {Chuvilin}}, \bibinfo {author} {\bibfnamefont {I.}~\bibnamefont {Mertig}}, \bibinfo {author} {\bibfnamefont {M.}~\bibnamefont {Gobbi}}, \bibinfo {author} {\bibfnamefont {F.}~\bibnamefont {Casanova}},\ and\ \bibinfo {author} {\bibfnamefont {L.~E.}\ \bibnamefont {Hueso}},\ }\href@noop {} {\bibfield  {journal} {\bibinfo {title} {Gate-tuneable and chirality-dependent charge-to-spin conversion in tellurium nanowires.}} {\bibinfo  {journal} {Nat. Mater.}\
  }\textbf {\bibinfo {volume} {21}},\ \bibinfo {pages} {526} (\bibinfo {year} {2022})}\BibitemShut {NoStop}%
\bibitem [{\citenamefont {Hirobe}\ \emph {et~al.}(2022)\citenamefont {Hirobe}, \citenamefont {Nabei},\ and\ \citenamefont {Yamamoto}}]{hirobe2022chirality}%
  \BibitemOpen
  \bibfield  {author} {\bibinfo {author} {\bibfnamefont {D.}~\bibnamefont {Hirobe}}, \bibinfo {author} {\bibfnamefont {Y.}~\bibnamefont {Nabei}},\ and\ \bibinfo {author} {\bibfnamefont {H.~M.}\ \bibnamefont {Yamamoto}},\ }\href@noop {} {\bibfield  {journal} {\bibinfo {title} {Chirality-induced intrinsic charge rectification in a tellurium-based field-effect transistor.}} {\bibinfo  {journal} {Phys. Rev. B}\ }\textbf {\bibinfo {volume} {106}},\ \bibinfo {pages} {L220403} (\bibinfo {year} {2022})}\BibitemShut {NoStop}%
\bibitem [{\citenamefont {Sudo}\ \emph {et~al.}(2023)\citenamefont {Sudo}, \citenamefont {Yanagi}, \citenamefont {Takahashi}, \citenamefont {Huynh}, \citenamefont {Tanigaki}, \citenamefont {Kobayashi}, \citenamefont {Suzuki},\ and\ \citenamefont {Kimata}}]{sudo2023valley}%
  \BibitemOpen
  \bibfield  {author} {\bibinfo {author} {\bibfnamefont {K.}~\bibnamefont {Sudo}}, \bibinfo {author} {\bibfnamefont {Y.}~\bibnamefont {Yanagi}}, \bibinfo {author} {\bibfnamefont {T.}~\bibnamefont {Takahashi}}, \bibinfo {author} {\bibfnamefont {K.-K.}\ \bibnamefont {Huynh}}, \bibinfo {author} {\bibfnamefont {K.}~\bibnamefont {Tanigaki}}, \bibinfo {author} {\bibfnamefont {K.}~\bibnamefont {Kobayashi}}, \bibinfo {author} {\bibfnamefont {M.-T.}\ \bibnamefont {Suzuki}},\ and\ \bibinfo {author} {\bibfnamefont {M.}~\bibnamefont {Kimata}},\ }\href@noop {} {\bibfield  {journal} {\bibinfo {title} {Valley polarization dependence of nonreciprocal transport in a chiral semiconductor.}} {\bibinfo  {journal} {Phys. Rev. B}\ }\textbf {\bibinfo {volume} {108}},\ \bibinfo {pages} {125137} (\bibinfo {year} {2023})}\BibitemShut {NoStop}%
\bibitem [{\citenamefont {Niu}\ \emph {et~al.}(2023{\natexlab{a}})\citenamefont {Niu}, \citenamefont {Qiu}, \citenamefont {Wang}, \citenamefont {Tan}, \citenamefont {Wang}, \citenamefont {Jian}, \citenamefont {Wang}, \citenamefont {Wu},\ and\ \citenamefont {Ye}}]{niu2023tunable00}%
  \BibitemOpen
  \bibfield  {author} {\bibinfo {author} {\bibfnamefont {C.}~\bibnamefont {Niu}}, \bibinfo {author} {\bibfnamefont {G.}~\bibnamefont {Qiu}}, \bibinfo {author} {\bibfnamefont {Y.}~\bibnamefont {Wang}}, \bibinfo {author} {\bibfnamefont {P.}~\bibnamefont {Tan}}, \bibinfo {author} {\bibfnamefont {M.}~\bibnamefont {Wang}}, \bibinfo {author} {\bibfnamefont {J.}~\bibnamefont {Jian}}, \bibinfo {author} {\bibfnamefont {H.}~\bibnamefont {Wang}}, \bibinfo {author} {\bibfnamefont {W.}~\bibnamefont {Wu}},\ and\ \bibinfo {author} {\bibfnamefont {P.~D.}\ \bibnamefont {Ye}},\ }\href@noop {} {\bibfield  {journal} {\bibinfo {title} {Tunable Chirality-Dependent Nonlinear Electrical Responses in 2D Tellurium.}} {\bibinfo  {journal} {Nano Lett.}\ }\textbf {\bibinfo {volume} {23}},\ \bibinfo {pages} {8445} (\bibinfo {year} {2023}{\natexlab{a}})}\BibitemShut {NoStop}%
\bibitem [{\citenamefont {Cheng}\ \emph {et~al.}(2019)\citenamefont {Cheng}, \citenamefont {Wu}, \citenamefont {Zhu},\ and\ \citenamefont {Guo}}]{ChengPRB2019}%
  \BibitemOpen
  \bibfield  {author} {\bibinfo {author} {\bibfnamefont {M.}~\bibnamefont {Cheng}}, \bibinfo {author} {\bibfnamefont {S.}~\bibnamefont {Wu}}, \bibinfo {author} {\bibfnamefont {Z.-Z.}\ \bibnamefont {Zhu}},\ and\ \bibinfo {author} {\bibfnamefont {G.-Y.}\ \bibnamefont {Guo}},\ }\href@noop {} {\bibfield  {journal} {\bibinfo {title} {Large secondharmonic generation and linear electro-optic effect in trigonal selenium and tellurium.}} {\bibinfo  {journal} {Phys. Rev. B}\ }\textbf {\bibinfo {volume} {100}},\ \bibinfo {pages} {035202} (\bibinfo {year} {2019})}\BibitemShut {NoStop}%
\bibitem [{\citenamefont {Fu}\ \emph {et~al.}(2023)\citenamefont {Fu}, \citenamefont {Cong}, \citenamefont {Xu}, \citenamefont {Zhu}, \citenamefont {Zhao}, \citenamefont {Liu}, \citenamefont {Yao}, \citenamefont {Xu}, \citenamefont {Deng}, \citenamefont {Zhu} \emph {et~al.}}]{FuAdvMater2023}%
  \BibitemOpen
  \bibfield  {author} {\bibinfo {author} {\bibfnamefont {Q.}~\bibnamefont {Fu}}, \bibinfo {author} {\bibfnamefont {X.}~\bibnamefont {Cong}}, \bibinfo {author} {\bibfnamefont {X.}~\bibnamefont {Xu}}, \bibinfo {author} {\bibfnamefont {S.}~\bibnamefont {Zhu}}, \bibinfo {author} {\bibfnamefont {X.}~\bibnamefont {Zhao}}, \bibinfo {author} {\bibfnamefont {S.}~\bibnamefont {Liu}}, \bibinfo {author} {\bibfnamefont {B.}~\bibnamefont {Yao}}, \bibinfo {author} {\bibfnamefont {M.}~\bibnamefont {Xu}}, \bibinfo {author} {\bibfnamefont {Y.}~\bibnamefont {Deng}}, \bibinfo {author} {\bibfnamefont {C.}~\bibnamefont {Zhu}}, \emph {et~al.},\ }\href@noop {} {\bibfield  {journal} {\bibinfo {title} {Berry Curvature Dipole Induced Giant Mid-Infrared Second-Harmonic Generation in 2D Weyl Semiconductor.}} {\bibinfo  {journal} {Adv. Mater.}\ }\textbf {\bibinfo {volume} {35}},\ \bibinfo {pages} {2306330} (\bibinfo {year} {2023})}\BibitemShut {NoStop}%
\bibitem [{\citenamefont {Niu}\ \emph {et~al.}(2023{\natexlab{b}})\citenamefont {Niu}, \citenamefont {Huang}, \citenamefont {Ghosh}, \citenamefont {Tan}, \citenamefont {Wang}, \citenamefont {Wu}, \citenamefont {Xu},\ and\ \citenamefont {Ye}}]{niu2023tunable01}%
  \BibitemOpen
  \bibfield  {author} {\bibinfo {author} {\bibfnamefont {C.}~\bibnamefont {Niu}}, \bibinfo {author} {\bibfnamefont {S.}~\bibnamefont {Huang}}, \bibinfo {author} {\bibfnamefont {N.}~\bibnamefont {Ghosh}}, \bibinfo {author} {\bibfnamefont {P.}~\bibnamefont {Tan}}, \bibinfo {author} {\bibfnamefont {M.}~\bibnamefont {Wang}}, \bibinfo {author} {\bibfnamefont {W.}~\bibnamefont {Wu}}, \bibinfo {author} {\bibfnamefont {X.}~\bibnamefont {Xu}},\ and\ \bibinfo {author} {\bibfnamefont {P.~D.}\ \bibnamefont {Ye}},\ }\href@noop {} {\bibfield  {journal} {\bibinfo {title} {Tunable circular photogalvanic and photovoltaic effect in 2D tellurium with different chirality.}} {\bibinfo  {journal} {Nano Lett.}\ }\textbf {\bibinfo {volume} {23}},\ \bibinfo {pages} {3599} (\bibinfo {year} {2023}{\natexlab{b}})}\BibitemShut {NoStop}%
\bibitem [{\citenamefont {Tanaka}\ \emph {et~al.}(2010)\citenamefont {Tanaka}, \citenamefont {Collins}, \citenamefont {Lovesey}, \citenamefont {Matsumami}, \citenamefont {Moriwaki},\ and\ \citenamefont {Shin}}]{TanakaJPCM2010}%
  \BibitemOpen
  \bibfield  {author} {\bibinfo {author} {\bibfnamefont {Y.}~\bibnamefont {Tanaka}}, \bibinfo {author} {\bibfnamefont {S.~P.}\ \bibnamefont {Collins}}, \bibinfo {author} {\bibfnamefont {S.~W.}\ \bibnamefont {Lovesey}}, \bibinfo {author} {\bibfnamefont {M.}~\bibnamefont {Matsumami}}, \bibinfo {author} {\bibfnamefont {T.}~\bibnamefont {Moriwaki}},\ and\ \bibinfo {author} {\bibfnamefont {S.}~\bibnamefont {Shin}},\ }\href@noop {} {\bibfield  {journal} {\bibinfo {title} {Determination of the absolute chirality of tellurium using resonant diffraction with circularly polarized x-rays.}} {\bibinfo  {journal} {J. Phys.: Condens. Matter}\ }\textbf {\bibinfo {volume} {22}},\ \bibinfo {pages} {122201} (\bibinfo {year} {2010})}\BibitemShut {NoStop}%
\bibitem [{\citenamefont {Betbeder-Matibet}\ and\ \citenamefont {Hulin}(1969)}]{MatibetPSS1969}%
  \BibitemOpen
  \bibfield  {author} {\bibinfo {author} {\bibfnamefont {O.}~\bibnamefont {Betbeder-Matibet}}\ and\ \bibinfo {author} {\bibfnamefont {M.}~\bibnamefont {Hulin}},\ }\href@noop {} {\bibfield  {journal} {\bibinfo {title} {A. Semi-Empirical Model for the Valence Band Structure of Tellurium.}} {\bibinfo  {journal} {Phys. Stat. Sol.}\ }\textbf {\bibinfo {volume} {36}},\ \bibinfo {pages} {573} (\bibinfo {year} {1969})}\BibitemShut {NoStop}%
\bibitem [{\citenamefont {Doi}\ \emph {et~al.}(1970)\citenamefont {Doi}, \citenamefont {K.Nakao},\ and\ \citenamefont {H.Kamimura}}]{DoiJPSJ1970}%
  \BibitemOpen
  \bibfield  {author} {\bibinfo {author} {\bibfnamefont {T.}~\bibnamefont {Doi}}, \bibinfo {author} {\bibnamefont {K.Nakao}},\ and\ \bibinfo {author} {\bibnamefont {H.Kamimura}},\ }\href@noop {} {\bibfield  {journal} {\bibinfo {title} {The Valence Band Structure of Tellurium. I. The $k \cdot p$ Perturbation Method.}} {\bibinfo  {journal} {J. Phys. Soc. Jpn.}\ }\textbf {\bibinfo {volume} {28}},\ \bibinfo {pages} {36} (\bibinfo {year} {1970})}\BibitemShut {NoStop}%
\bibitem [{\citenamefont {Joannopoulos}\ \emph {et~al.}(1975)\citenamefont {Joannopoulos}, \citenamefont {Schl{\"u}ter},\ and\ \citenamefont {Cohen}}]{joannopoulos1975electronic}%
  \BibitemOpen
  \bibfield  {author} {\bibinfo {author} {\bibfnamefont {J.}~\bibnamefont {Joannopoulos}}, \bibinfo {author} {\bibfnamefont {M.}~\bibnamefont {Schl{\"u}ter}},\ and\ \bibinfo {author} {\bibfnamefont {M.~L.}\ \bibnamefont {Cohen}},\ }\href@noop {} {\bibfield  {journal} {\bibinfo {title} {Electronic structure of trigonal and amorphous Se and Te.}} {\bibinfo  {journal} {Phys. Rev. B}\ }\textbf {\bibinfo {volume} {11}},\ \bibinfo {pages} {2186} (\bibinfo {year} {1975})}\BibitemShut {NoStop}%
\bibitem [{\citenamefont {Asendorf}(1957)}]{asendorf1957space}%
  \BibitemOpen
  \bibfield  {author} {\bibinfo {author} {\bibfnamefont {R.~H.}\ \bibnamefont {Asendorf}},\ }\href@noop {} {\bibfield  {journal} {\bibinfo {title} {Space group of tellurium and selenium.}} {\bibinfo  {journal} {J. Chem. Phys.}\ }\textbf {\bibinfo {volume} {27}},\ \bibinfo {pages} {11} (\bibinfo {year} {1957})}\BibitemShut {NoStop}%
\bibitem [{\citenamefont {Hirayama}\ \emph {et~al.}(2015)\citenamefont {Hirayama}, \citenamefont {Okugawa}, \citenamefont {Ishibashi}, \citenamefont {Murakami},\ and\ \citenamefont {Miyake}}]{HirayamaPRL2015}%
  \BibitemOpen
  \bibfield  {author} {\bibinfo {author} {\bibfnamefont {M.}~\bibnamefont {Hirayama}}, \bibinfo {author} {\bibfnamefont {R.}~\bibnamefont {Okugawa}}, \bibinfo {author} {\bibfnamefont {S.}~\bibnamefont {Ishibashi}}, \bibinfo {author} {\bibfnamefont {S.}~\bibnamefont {Murakami}},\ and\ \bibinfo {author} {\bibfnamefont {T.}~\bibnamefont {Miyake}},\ }\href@noop {} {\bibfield  {journal} {\bibinfo {title} {Weyl Node and Spin Texture in Trigonal Tellurium and Selenium.}} {\bibinfo  {journal} {Phys. Rev. Lett.}\ }\textbf {\bibinfo {volume} {114}},\ \bibinfo {pages} {206401} (\bibinfo {year} {2015})}\BibitemShut {NoStop}%
\bibitem [{\citenamefont {Peng}\ \emph {et~al.}(2014)\citenamefont {Peng}, \citenamefont {Kioussis},\ and\ \citenamefont {Snyder}}]{peng2014elemental}%
  \BibitemOpen
  \bibfield  {author} {\bibinfo {author} {\bibfnamefont {H.}~\bibnamefont {Peng}}, \bibinfo {author} {\bibfnamefont {N.}~\bibnamefont {Kioussis}},\ and\ \bibinfo {author} {\bibfnamefont {G.~J.}\ \bibnamefont {Snyder}},\ }\href@noop {} {\bibfield  {journal} {\bibinfo {title} {Elemental tellurium as a chiral p-type thermoelectric material.}} {\bibinfo  {journal} {Phys. Rev. B}\ }\textbf {\bibinfo {volume} {89}},\ \bibinfo {pages} {195206} (\bibinfo {year} {2014})}\BibitemShut {NoStop}%
\bibitem [{\citenamefont {Ivchenko}\ and\ \citenamefont {Pikus}(1978)}]{ivchenko1978new}%
  \BibitemOpen
  \bibfield  {author} {\bibinfo {author} {\bibfnamefont {E.~L.}\ \bibnamefont {Ivchenko}}\ and\ \bibinfo {author} {\bibfnamefont {G.~E.}\ \bibnamefont {Pikus}},\ }\href@noop {} {\bibfield  {journal} {\bibinfo {title} {New photogalvanic effect in gyrotropic crystals.}} {\bibinfo  {journal} {JETP Lett.}\ }\textbf {\bibinfo {volume} {27}},\ \bibinfo {pages} {604} (\bibinfo {year} {1978})}\BibitemShut {NoStop}%
\bibitem [{\citenamefont {Tsirkin}\ \emph {et~al.}(2018)\citenamefont {Tsirkin}, \citenamefont {Puente},\ and\ \citenamefont {Souza}}]{TsirkinPRB2018}%
  \BibitemOpen
  \bibfield  {author} {\bibinfo {author} {\bibfnamefont {S.~S.}\ \bibnamefont {Tsirkin}}, \bibinfo {author} {\bibfnamefont {P.~A.}\ \bibnamefont {Puente}},\ and\ \bibinfo {author} {\bibfnamefont {I.}~\bibnamefont {Souza}},\ }\href@noop {} {\bibfield  {journal} {\bibinfo {title} {Gyrotropic effects in trigonal tellurium studied from first principles.}} {\bibinfo  {journal} {Phys. Rev. B}\ }\textbf {\bibinfo {volume} {97}},\ \bibinfo {pages} {035158} (\bibinfo {year} {2018})}\BibitemShut {NoStop}%
\bibitem [{\citenamefont {{\c{S}}ahin}\ \emph {et~al.}(2018)\citenamefont {{\c{S}}ahin}, \citenamefont {Rou}, \citenamefont {Ma},\ and\ \citenamefont {Pesin}}]{csahin2018pancharatnam}%
  \BibitemOpen
  \bibfield  {author} {\bibinfo {author} {\bibfnamefont {C.}~\bibnamefont {{\c{S}}ahin}}, \bibinfo {author} {\bibfnamefont {J.}~\bibnamefont {Rou}}, \bibinfo {author} {\bibfnamefont {J.}~\bibnamefont {Ma}},\ and\ \bibinfo {author} {\bibfnamefont {D.}~\bibnamefont {Pesin}},\ }\href@noop {} {\bibfield  {journal} {\bibinfo {title} {Pancharatnam-Berry phase and kinetic magnetoelectric effect in trigonal tellurium.}} {\bibinfo  {journal} {Phys. Rev. B}\ }\textbf {\bibinfo {volume} {97}},\ \bibinfo {pages} {205206} (\bibinfo {year} {2018})}\BibitemShut {NoStop}%
\bibitem [{SM()}]{SM}%
  \BibitemOpen
  \href@noop {} {}\bibinfo {note} {See Supplemental Material at [URL will be inserted by publisher] for parameters used in the tight-binding model; details of hydrothermal synthesis and lamination of Te crystals; details of the etching method and scanning electron micrsocpe images of etch pits over wide area; temperature-variable harmonic resistances of devices A and B; comparison between four- and two-terminal measurements of nonlinear electrical conduction; expression for the normalized second harmonic resistance in terms of linear and nonlinear electrical conductances; double-checking of the sign of the nonlinear electical conductance by d.c. magnetoconductance measurement; calculation of the nonlinear electrical conductivity by Boltzmann kinetic equation; and comparison of calculated carrier-density dependences of the nonlinear conductances due to orbital and spin magnetic moments, which are reconstructed from the inset to Fig.~4(d). The Supplemental Material also contains Refs.~\cite{schliemann2003anisotropic, vyborny2009semiclassical, liu2016mobility, xiao2016unconventional, kim2019vertex}.}\BibitemShut {Stop}%
\bibitem [{S1()}]{schliemann2003anisotropic}%
  \BibitemOpen
  \bibfield  {author} {\bibinfo {author} {\bibfnamefont {J.}~\bibnamefont {Schliemann}}\ and\ \bibinfo {author} {\bibfnamefont {D.}~\bibnamefont {Loss}},\ }\href@noop {} {\bibfield  {journal} {\bibinfo {title} {Anisotropic transport in a two-dimensional electron gas in the presence of spin-orbit coupling.}} {\bibinfo  {journal} {Phys. Rev. B}\ }\textbf {\bibinfo {volume} {68}},\ \bibinfo {pages} {165311} (\bibinfo {year} {2003})}\BibitemShut {NoStop}%
\bibitem [{S2()}]{vyborny2009semiclassical}%
  \BibitemOpen
  \bibfield  {author} {\bibinfo {author} {\bibfnamefont {K.}~\bibnamefont {V{\`y}born{\`y}}}, \bibinfo {author} {\bibfnamefont {A.~A.}\ \bibnamefont {Kovalev}}, \bibinfo {author} {\bibfnamefont {J.}~\bibnamefont {Sinova}},\ and\ \bibinfo {author} {\bibfnamefont {T.}~\bibnamefont {Jungwirth}},\ }\href@noop {} {\bibfield  {journal} {\bibinfo {title} {Semiclassical framework for the calculation of transport anisotropies.}} {\bibinfo  {journal} {Phys. Rev. B}\ }\textbf {\bibinfo {volume} {79}},\ \bibinfo {pages} {045427} (\bibinfo {year} {2009})}\BibitemShut {NoStop}%
\bibitem [{S3()}]{liu2016mobility}%
  \BibitemOpen
  \bibfield  {author} {\bibinfo {author} {\bibfnamefont {Y.}~\bibnamefont {Liu}}, \bibinfo {author} {\bibfnamefont {T.}~\bibnamefont {Low}},\ and\ \bibinfo {author} {\bibfnamefont {P.~P.}\ \bibnamefont {Ruden}},\ }\href@noop {} {\bibfield  {journal} {\bibinfo {title} {Mobility anisotropy in monolayer black phosphorus due to scattering by charged impurities.}} {\bibinfo  {journal} {Phys. Rev. B}\ }\textbf {\bibinfo {volume} {93}},\ \bibinfo {pages} {165402} (\bibinfo {year} {2016})}\BibitemShut {NoStop}%
\bibitem [{S4()}]{xiao2016unconventional}%
  \BibitemOpen
  \bibfield  {author} {\bibinfo {author} {\bibfnamefont {C.}~\bibnamefont {Xiao}}, \bibinfo {author} {\bibfnamefont {D.}~\bibnamefont {Li}},\ and\ \bibinfo {author} {\bibfnamefont {Z.}~\bibnamefont {Ma}},\ }\href@noop {} {\bibfield  {journal} {bibinfo {title} {Unconventional thermoelectric behaviors and enhancement of figure of merit in Rashba spintronic systems.}} {\bibinfo  {journal} {Phys. Rev. B}\ }\textbf {\bibinfo {volume} {93}},\ \bibinfo {pages} {075150} (\bibinfo {year} {2016})}\BibitemShut {NoStop}%
\bibitem [{S5()}]{kim2019vertex}%
  \BibitemOpen
  \bibfield  {author} {\bibinfo {author} {\bibfnamefont {S.}~\bibnamefont {Kim}}, \bibinfo {author} {\bibfnamefont {S.}~\bibnamefont {Woo}},\ and\ \bibinfo {author} {\bibfnamefont {H.}~\bibnamefont {Min}},\ }\href@noop {} {\bibfield  {journal} {\bibinfo {title} {Vertex corrections to the dc conductivity in anisotropic multiband systems.}} {\bibinfo  {journal} {Phys. Rev. B}\ }\textbf {\bibinfo {volume} {99}},\ \bibinfo {pages} {165107} (\bibinfo {year} {2019})}\BibitemShut {NoStop}%
\bibitem [{\citenamefont {Wang}\ \emph {et~al.}(2018)\citenamefont {Wang}, \citenamefont {Qiu}, \citenamefont {Wang}, \citenamefont {Huang}, \citenamefont {Wang}, \citenamefont {Liu}, \citenamefont {Du}, \citenamefont {III}, \citenamefont {Kim}, \citenamefont {Xu}, \citenamefont {Ye},\ and\ \citenamefont {Wu}}]{WanNatElectron2018}%
  \BibitemOpen
  \bibfield  {author} {\bibinfo {author} {\bibfnamefont {Y.}~\bibnamefont {Wang}}, \bibinfo {author} {\bibfnamefont {G.}~\bibnamefont {Qiu}}, \bibinfo {author} {\bibfnamefont {R.}~\bibnamefont {Wang}}, \bibinfo {author} {\bibfnamefont {S.}~\bibnamefont {Huang}}, \bibinfo {author} {\bibfnamefont {Q.}~\bibnamefont {Wang}}, \bibinfo {author} {\bibfnamefont {Y.}~\bibnamefont {Liu}}, \bibinfo {author} {\bibfnamefont {Y.}~\bibnamefont {Du}}, \bibinfo {author} {\bibfnamefont {W.~A.~G.}\ \bibnamefont {III}}, \bibinfo {author} {\bibfnamefont {M.~J.}\ \bibnamefont {Kim}}, \bibinfo {author} {\bibfnamefont {X.}~\bibnamefont {Xu}}, \bibinfo {author} {\bibfnamefont {P.~D.}\ \bibnamefont {Ye}},\ and\ \bibinfo {author} {\bibfnamefont {W.}~\bibnamefont {Wu}},\ }\href@noop {} {\bibfield  {journal} {\bibinfo {title} {Field-effect transistors made from solution-grown two-dimensional tellurene.}} {\bibinfo  {journal} {Nat. Electron.}\ }\textbf {\bibinfo {volume} {1}},\ \bibinfo {pages} {228} (\bibinfo {year} {2018})}\BibitemShut {NoStop}%
\bibitem [{\citenamefont {Qiu}\ \emph {et~al.}(2019)\citenamefont {Qiu}, \citenamefont {Huang}, \citenamefont {Segovia}, \citenamefont {Venuthurumilli}, \citenamefont {Wang}, \citenamefont {Wu}, \citenamefont {Xu},\ and\ \citenamefont {Ye}}]{qiu2019thermoelectric}%
  \BibitemOpen
  \bibfield  {author} {\bibinfo {author} {\bibfnamefont {G.}~\bibnamefont {Qiu}}, \bibinfo {author} {\bibfnamefont {S.}~\bibnamefont {Huang}}, \bibinfo {author} {\bibfnamefont {M.}~\bibnamefont {Segovia}}, \bibinfo {author} {\bibfnamefont {P.~K.}\ \bibnamefont {Venuthurumilli}}, \bibinfo {author} {\bibfnamefont {Y.}~\bibnamefont {Wang}}, \bibinfo {author} {\bibfnamefont {W.}~\bibnamefont {Wu}}, \bibinfo {author} {\bibfnamefont {X.}~\bibnamefont {Xu}},\ and\ \bibinfo {author} {\bibfnamefont {P.~D.}\ \bibnamefont {Ye}},\ }\href@noop {} {\bibfield  {journal} {\bibinfo {title} {Thermoelectric performance of 2D tellurium with accumulation contacts.}} {\bibinfo  {journal} {Nano Lett.}\ }\textbf {\bibinfo {volume} {19}},\ \bibinfo {pages} {1955} (\bibinfo {year} {2019})}\BibitemShut {NoStop}%
\bibitem [{\citenamefont {Golub}\ \emph {et~al.}(2023)\citenamefont {Golub}, \citenamefont {Ivchenko},\ and\ \citenamefont {Spivak}}]{golub2023electrical}%
  \BibitemOpen
  \bibfield  {author} {\bibinfo {author} {\bibfnamefont {L.}~\bibnamefont {Golub}}, \bibinfo {author} {\bibfnamefont {E.}~\bibnamefont {Ivchenko}},\ and\ \bibinfo {author} {\bibfnamefont {B.}~\bibnamefont {Spivak}},\ }\href@noop {} {\bibfield  {journal} {\bibinfo {title} {Electrical magnetochiral current in tellurium.}} {\bibinfo  {journal} {Phys. Rev. B}\ }\textbf {\bibinfo {volume} {108}},\ \bibinfo {pages} {245202} (\bibinfo {year} {2023})}\BibitemShut {NoStop}%
\bibitem [{\citenamefont {Nakazawa}\ \emph {et~al.}(2024)\citenamefont {Nakazawa}, \citenamefont {Yamaguchi},\ and\ \citenamefont {Yamakage}}]{nakazawa2024nonlinear}%
  \BibitemOpen
  \bibfield  {author} {\bibinfo {author} {\bibfnamefont {K.}~\bibnamefont {Nakazawa}}, \bibinfo {author} {\bibfnamefont {T.}~\bibnamefont {Yamaguchi}},\ and\ \bibinfo {author} {\bibfnamefont {A.}~\bibnamefont {Yamakage}},\ }\Eprint {https://arxiv.org/abs/2403.10337} {Nonlinear charge transport properties in chiral tellurium.} {arXiv:2403.10337 [cond-mat.str-el]}  (\bibinfo {year} {2024})\BibitemShut {NoStop}%
\end{thebibliography}
\end{document}